\documentclass{jfm}
\usepackage{graphicx}
\usepackage{natbib}
\usepackage[usenames,dvipsnames]{color}

\ifCUPmtlplainloaded \else
  \checkfont{eurm10}
  \iffontfound
    \IfFileExists{upmath.sty}
      {\typeout{^^JFound AMS Euler Roman fonts on the system,
                   using the 'upmath' package.^^J}%
       \usepackage{upmath}}
      {\typeout{^^JFound AMS Euler Roman fonts on the system, but you
                   dont seem to have the}%
       \typeout{'upmath' package installed. JFM.cls can take advantage
                 of these fonts,^^Jif you use 'upmath' package.^^J}%
      }
  \else
  \fi
\fi

\ifCUPmtlplainloaded \else
  \checkfont{msam10}
  \iffontfound
    \IfFileExists{amssymb.sty}
      {\typeout{^^JFound AMS Symbol fonts on the system, using the
                'amssymb' package.^^J}%
       \usepackage{amssymb}%
         \let\leq=\leqslant
         \let\geq=\geqslant
      }{}
  \fi
\fi

\ifCUPmtlplainloaded \else
  \IfFileExists{amsbsy.sty}
    {\typeout{^^JFound the 'amsbsy' package on the system, using it.^^J}%
     \usepackage{amsbsy}}
    {\providecommand\boldsymbol[1]{\mbox{\boldmath $##1$}}}
\fi

\newcommand\Rey{\mbox{\textit{Re}}}  % Reynolds number
\newsavebox{\astrutbox}
\sbox{\astrutbox}{\rule[-5pt]{0pt}{20pt}}

\def\lx{\lambda_x}
\def\lz{\lambda_z}

\def\LSMBOX{{LSM-Box}}

\def\VLSMBOX{{VLSM-Box}}

        \def\bfU{\mathbf{U}}
        \def\bfu{\mathbf{u}}
        
        \def\reffig#1{figure~\ref{fig:#1}}
        
        \def\refsec#1{\S\ref{sec:#1}}
        \def\reftab#1{table~\ref{tab:#1}}
        \def\Rey{{\rm Re}}
        \def\Retau{{\rm Re}_\tau}
        \def\vel      {\bfu}
        \def\basevel  {\bfU}

        \def\peanuts{{\tt peanuts}}
       \def\tws{steady~}

\title[Self-sustained large-scale motions in turbulent Couette flow]{On the self-sustained nature of large-scale motions in turbulent Couette flow}

\author[S. Rawat, C. Cossu, Y. Hwang, and F. Rincon ]%
 {S\ls U\ls B\ls H\ls A\ls N\ls D\ls U \ns R\ls A\ls W\ls A\ls T$^{1}$,
 \ns C\ls A\ls R\ls L\ls O\ns  C\ls O\ls S\ls S\ls U$^{1}$\thanks{Email address for correspondence: carlo.cossu@imft.fr},
Y\ls O\ls N\ls G\ls Y\ls U\ls N\ls \ns H\ls W\ls A\ls N\ls G$^{2}$
 \and \ns F\ls R\ls A\ls N\ls \c C\ls O\ls I\ls S\ns  R\ls I\ls N\ls C\ls O\ls N$^{3,4}$}
\affiliation{$^1$ Institut de M\'ecanique des Fluides de Toulouse (IMFT), CNRS -- Universit\'e de Toulouse\\
All\'ee du Pr. Camille Soula, F-31400 Toulouse, France\\[\affilskip]
$^2$ Department of Aeronautics, Imperial College, South Kensington, London SW7 2AZ, UK
\\[\affilskip]
$^3$Universit\'e de Toulouse; UPS-OMP; IRAP; Toulouse, France\\[\affilskip]
$^4$CNRS; IRAP; 14 avenue Edouard Belin, F-31400 Toulouse, France
}

\date{}
\begin{document}

\maketitle

% {\bf REVISION: \add{ADDED} / \del{REMOVED} / \mod{MODIFIED}}

\begin{abstract}
Large-scale motions in wall-bounded turbulent flows are frequently interpreted as resulting from an aggregation process of smaller-scale structures.
Here, we explore the alternative possibility that such large-scale motions are themselves
self-sustained and do not draw their energy from smaller-scale turbulent motions activated in buffer layers.
To this end, it is first shown that large-scale motions in turbulent Couette flow at $\Rey=2150$ self-sustain even when active processes at smaller scales are artificially quenched by increasing the Smagorinsky constant $C_s$ in large eddy simulations. 
These results are in agreement with earlier results on pressure driven turbulent channels.
We further investigate the nature of the large-scale coherent motions by computing upper and lower-branch nonlinear steady solutions of the filtered (LES) equations with a Newton-Krylov solver, and find that they are connected by a saddle-node bifurcation at large values of $C_s$.  
Upper branch solutions for the filtered large scale motions are computed for Reynolds numbers up to $\Rey=2187$ using specific paths in the $\Rey-C_s$ parameter plane and compared to large-scale coherent motions.
Continuation to $C_s=0$ reveals that these large-scale steady solutions of the filtered equations are connected to the Nagata-Clever-Busse-Waleffe branch of steady solutions of the Navier-Stokes equations.
 In contrast, we find it impossible to connect the latter to buffer layer motions through a continuation to higher Reynolds numbers in minimal flow units.
\end{abstract}

%%%%%%%%%%%%%%%%%%%%%%%%%%%%%%%%%%%%%%%%%%%%%%%%%%%%%%%%%%%%%%%%%%%%%%%%%%%%%%%
%%%%%%%%%%%%%%%%%%%%%%%%%%%%%%%%%%%%%%%%%%%%%%%%%%%%%%%%%%%%%%%%%%%%%%%%%%%%%%%
\section{Introduction}

%...............................................................................................
\subsection{Streaky motions in wall-bounded turbulent flows}

One of the most robust features observed in wall-bounded turbulent shear flows is the presence of quasi-streamwise streaks, i.e. narrow streamwise regions where the streamwise velocity is larger or smaller than its mean value.
The flow visualizations of \cite{Kline1967} revealed that the near-wall region of turbulent boundary layers  is very active and is populated by streamwise streaks in a spanwise quasi-periodic pattern.
The average spanwise streak-spacing in the viscous sublayer and the buffer layer scales in wall units and corresponds approximately to $\lambda_z^+ = 100$ \citep{Kline1967,Smith1983}.
This spacing has been confirmed by the first direct numerical simulation of channel flow at $Re_{\tau} \approx 180$ by \cite{Kim1987}, which also revealed the existence of counter-rotating quasi-streamwise vortices.

Streaky motions also exist at much larger scales.
It has long been known that in turbulent flows the outer region is dominated by large-scale motions (LSM) with dimensions of the order of the outer length scale  $h$ (e.g. the channel half-width or the boundary layer $\delta_{99}$ thickness) often  separated by regions of non-turbulent fluid \citep{Corrsin1954,Kovasznay1970,Blackwelder1972}. 
More recently it has been realized that in addition to large-scale motions, `very large-scale motions' (VLSM) exist with streamwise extensions at least fourfold those of the LSM.
Large and very-large scale motions have been observed in turbulent
pipe flow \cite[]{Kim1999}, turbulent plane Poiseuille flow
\cite[]{delAlamo2003} and in the turbulent boundary layer \cite[]{Tomkins2003,Tomkins2005,Hutchins2007}. 
At high Reynolds numbers, these structures dominate the streamwise turbulent kinetic energy throughout the logarithmic region \cite[]{Tomkins2003,Tomkins2005}, and modulate the cycles of near-wall structures \cite[]{Hutchins2007b,Mathis2009}.

As far as turbulent Couette flow is concerned, \cite{Lee1991} found large-scale motions in the form of quasi-steady large-scale coherent streaks and rolls in direct numerical simulations at $\Rey=3000$ (where $\Rey$ is based on the channel half-width $h$ and half of the velocity difference between the walls). 
These structures, of spanwise wavelength $\lambda_z \approx 4 h$, occupied half of the spanwise size ($L_z=8.4 h$) of the computational domain and its whole streamwise extension ($L_x=12.6 h$). 
The existence of large-scale streaks was confirmed by the experiments of \cite{Tillmark1994}.
In order to understand whether the characteristics of these large-scale structures were affected by the size of the computational box, \cite{Komminaho1996} repeated the computations in a larger domain ($L_x=88 h$ and $L_z = 25.1 h$) at $\Rey=750$. 
They found large and very-large scale coherent streaks  with the same
spanwise wavelength ($\approx 4\,h$) as reported by \cite{Lee1991}, but more unsteady and extending more than $30 h$ in the streamwise direction. 
They also found that resolving these structures is important to obtain accurate turbulent statistics.  
\cite{Kitoh2005}, \cite{Tsukahara2006}, \cite{Tsukahara2007} and \cite{Kitoh2008} also found large and very-large scale streaks at higher Reynolds numbers with typical spanwise spacings $\lambda_z\approx 4.2 h - 5 h$ and streamwise lengths attaining $\lambda_x \approx 40 h - 60 h$ for very-large scale motions.

%-----------------------------------------------------------------------------------------------------
\subsection{Mechanisms sustaining streaky motions}

There is now a relatively large consensus that streaky motions in the near-wall region rely on a self-sustained process.
That the process is self-sustained was demonstrated by \cite{Jimenez1991} who found that near-wall turbulence can be sustained in spanwise and streamwise periodic numerical domains as small as $L_x^+ \times L_z^+ \approx (250 - 300) \times 100$,
showing that the motions in the viscous region are sustained independently from motions at larger scales.
\cite{Hamilton1995} and \cite{Waleffe1995} decomposed this nonlinear self-sustained process into three basic mechanisms: the lift-up effect which amplifies quasi-streamwise vortices into quasi-streamwise streaks, the amplification of sinuous modes supported by the streaks, and the streaks breakdown supporting the regeneration of the vortices.
The lift-up is associated with the redistribution of streamwise
momentum by counter-rotating, spanwise periodic quasi-streamwise
vortices immersed in a basic shear flow. 
This redistribution leads to the transient amplification of high-velocity and low-velocity streamwise streaks \citep{Moffatt1967,Ellingsen1975,Landahl1980,Landahl1990,Schmid2001}. 
When the streaks reach sufficiently large amplitudes they become unstable to secondary perturbations via an inflectional, typically sinuous, instability \citep{Waleffe1995,Reddy1998}.
As this secondary instability is subcritical, sinuous modes can also develop on top of streaks of amplitude smaller than the critical one  \citep{Schoppa2002,Cossu2011}.
The breakdown of streamwise streaks finally leads to the regeneration of streamwise vorticity via nonlinear mechanisms.
For the process to be self-sustained, the Reynolds number and the spanwise length of the box need to be large enough to allow for sufficient energy amplification by the lift-up effect and the streamwise length needs to be large enough to allow the secondary instability to be sufficiently amplified. 

There is less consensus on the origin of large-scale motions.
Early investigations showed that these motions contain a number of smaller scale structures which have been interpreted as hairpin vortices \citep{Falco1977,Head1981}.
Motions at large scales have therefore been interpreted as generated by the mutual vortical induction \citep{Zhou1999,Adrian2000} and merger and growth of hairpins ultimately originating from the near-wall region \citep{Tomkins2003,Adrian2007}.
The idea that large-scale motions originate from motions at smaller scales is often implied in structural models of the logarithmic layer \citep[][and following extensions]{Perry1982}.
As for very large-scale motions, it has been proposed that they result from the streamwise concatenation of large-scale motions \citep[see e.g.][]{Kim1999,Guala2006,Dennis2011b}.
From a different perspective, \cite{Toh2005} conjectured the existence of a co-supporting cycle where near-wall structures sustain large-scale structures in plane Poiseuille flow.
The common point of these theories is that large-scale streaky motions would not exist in the absence of the near-wall active cycle. 

The idea that near-wall active motions are ultimately necessary to form large-scale motions is however challenged by a growing number of recent results. 
For instance, it has been shown that large-scale outer motions are not significantly influenced by the change of the near-wall dynamics induced by wall-roughness \citep{Flores2006,Flores2007} which implies that they are not very sensitive to the near-wall cycle. 
Also, some studies cast doubts on the very idea that hairpin vortices are a prominent feature of high Reynolds number turbulence \citep[see e.g.][]{Jeong1997,Eitel2015}.
Indeed, an alternative explanation of the origin of large-scale motions is currently emerging, which conjectures that these motions are sustained by a mechanism similar to the self-sustained process (SSP) proposed by \cite{Hamilton1995}. This self-sustained process has been identified in transitional flows but \cite{Hamilton1995} argue that the near-wall streaky motions identified by \cite{Kline1967} are based on the same mechanism.
An essential ingredient of this emerging scenario is the existence of a `coherent lift-up effect', which would allow coherent large-scale motions to extract energy from the mean flow. 
That this coherent lift-up effect exists was shown by \cite{delAlamo2006},  \cite{Pujals2009} and \cite{Cossu2009}, who computed the optimal linear transient growth sustained by turbulent mean flow profiles using the eddy viscosity model of \cite{Reynolds1972}.
Following the same approach \cite{Hwang2010,Hwang2010c} and \cite{Willis2010}  showed that  turbulent Poiseuille, Hagen-Poiseuille and plane Couette flows all bear very large amplifications of harmonic and stochastic forcing.
These different investigations all show that coherent streamwise and quasi-streamwise streaks can be amplified starting from coherent streamwise and quasi-streamwise vortices.
The most amplified spanwise scales have been shown to be in good agreement with the spanwise spacing of large-scale streaks in the outer region and the associated  maximum amplifications have been shown to increase with the Reynolds number.
\cite{Pujals2010b} experimentally confirmed the existence of the spatially-coherent transient growth of artificially forced large-scale structures in the turbulent boundary  layer.

The existence of a robust coherent lift-up effect and of a secondary
instability of large-scale streaks \citep{Park2011} are strong indications that a self-sustained process might be at work at large scales in turbulent flows. 
The confirmation of the existence of this process was given by \cite{Hwang2010b}, who showed that large-scale motions can self-sustain even in the absence of smaller-scale processes active in the near-wall and logarithmic regions. 
In order to suppress the small-scale motions while preserving the dissipation associated with them, \cite{Hwang2010b} used a large eddy simulation (LES) filter without energy backscatter and artificially increased its cutoff characteristic length. 
In this way, they were able to show that when the near-wall motions are artificially quenched, motions with the usual scales of large-scale motions survive.
These conclusions were later extended by \cite{Hwang2011} to intermediate flow units characteristic of motions in the logarithmic layer \citep{Flores2010}.

%====================================================================
\subsection{Goal of the present study}

The scope of the present study is twofold. 
The first, most immediate, objective is to determine if large-scale
motions are self-sustained also in  turbulent plane Couette flow.
This would provide the first confirmation of the findings of \cite{Hwang2010b} in a flow other than the pressure driven channel flow. For the sake of exposing the objectives of our study, we anticipate the result that large-scale motions are indeed
also self-sustained in turbulent plane Couette flow.

As a second objective we would like to further investigate the nature of the process by which these large-scale motions do self-sustain by using methods borrowed from dynamical systems theory.
In particular, the computation and analysis of invariant solutions of the Navier-Stokes equations has led to important progress in the understanding of subcritical transition in shear flows. 
A number of non-trivial steady state solutions have for instance been computed in plane Couette flow \citep{Nagata1990,Clever1992,Clever1997,Waleffe1998,Waleffe2003,Gibson2008}.
These steady state solutions do typically represent `saddles' in phase space and appear in a saddle-node bifurcation at low Reynolds numbers with additional solutions appearing when the Reynolds number is increased.
Lower branch solutions are related to the laminar-turbulent transition boundary, while upper branch solutions display features consistent with the turbulent flow issued from the transition process.
Most of these saddle solutions have only a few unstable eigenvalues. 
The original hope was that turbulent solutions spend a significant time in the neighbourhood of the relevant saddles when approaching them near their stable manifold and before being ejected along the unstable manifold. 
This has been later shown not to be the case \citep{Kerswell2007,Schneider2007}.
The current hope is that turbulent statistics could be captured by expansions based on the properties of unstable periodic solutions \citep{Artuso1990,Kawahara2001}. 
However, to this date, attempts to prove the relevance of this approach have not been completely successful \citep[see e.g.][]{Chandler2013}, which calls into question the relevance of using such solutions to describe fully developed, higher Reynolds number turbulent regimes in which their number increases very rapidly and their spatial structures become increasingly complex.

A possible way to by-pass this problem is to model small-scale motions, in order to take only their averaged effect into account, and to concentrate on large-scale coherent motions which contain most of the energy at high Reynolds numbers. 
In the second part of this study, we therefore attempt to compute invariant solutions corresponding to coherent large-scale motions. 
In particular we will look for steady solutions of the {\it filtered} motions, i.e. solutions of the same LES equations used in the first part of the study to show that large-scale motions are self-sustained.
Many questions of great interest can be investigated through this
approach, several of which will be addressed in the paper. Do steady state
solutions of the filtered equations exist?
If yes, how are they related to the steady solutions computed for the
Navier-Stokes equations, mainly at transitional Reynolds numbers? Are
these Navier-Stokes solutions relevant in fully developed turbulent flows? Do they evolve into near-wall or large-scale structures when the Reynolds number is raised sufficiently for inner-outer scale separation to set in?

The paper is organized as follows.
After a brief introduction in \refsec{background} of the mathematical formulation and main numerical tools used in our study, we show in \refsec{SSPLSM} that large-scale and very-large scale motions survive the quenching of the near-wall cycle in very large and long domains.
Nonlinear steady state solutions of the filtered (LES) equations are then computed, analysed, and discussed in \refsec{FTW}.  
The main results of our work are finally summarized and discussed in \refsec{disc}. Note that two auxiliary results are presented in Appendix to simplify the presentation. 
Appendix~\ref{sec:LESLSMBOX} shows that large-scale motions are also self-sustained in the absence of potentially active very-large scale motions of finite wavelength. Appendix~\ref{sec:BULABOX} shows that NBCW  solutions do not represent near-wall structures if continued to large Reynolds numbers in a minimal flow unit.

%%%%%%%%%%%%%%%%%%%%%%%%%%%%%%%%%%%%%%%%%%%%%%%%%%%%%%%%%%%%%%%%%%%%%%%%%%%%%%%
%%%%%%%%%%%%%%%%%%%%%%%%%%%%%%%%%%%%%%%%%%%%%%%%%%%%%%%%%%%%%%%%%%%%%%%%%%%%%%%
\section{Background}
\label{sec:background}

%====================================================================
\subsection{Computing coherent (filtered) motions}
\label{sec:back_LES}

We consider the plane Couette flow of a viscous fluid of constant density $\rho$ and kinematic viscosity $\nu$ between two parallel plates located at $y=\pm h$. 
The streamwise, wall-normal and spanwise coordinates are denoted by $x$, $y$ and $z$ respectively.
The two plates move in opposite directions with velocity $(\pm U_w,0,0)$ and the Reynolds number is defined as $\Rey= h U_w /\nu$. 
We are mainly interested in the fully developed turbulent regime where averaged quantities are of interest.
The mean velocity profile $U(y)$ and the root-mean-square ($rms$) velocity fluctuations $u_{rms}(y)$, $v_{rms}(y)$, $w_{rms}(y)$ are obtained by averaging the instantaneous fields over horizontal $x-z$ planes and in time (after the extinction of transients).
The friction Reynolds number $\Retau=u_\tau h / \nu$ is based on the friction velocity $u_\tau=\sqrt{\tau_w/\rho}$, where $\tau_w/\rho=\nu dU/dy|_w$.
We denote by the $^+$ superscript variables expressed in wall-units, i.e. made dimensionless with respect to $\nu / u_\tau$ for lengths and $u_\tau$ for velocities. 
Length expressed in wall units can be obtained by multiplication by $\Retau$ of those expressed in terms of the outer length $h$, so that, for instance, 
$y^+ = \Retau y /h$.

%LES: THE STATIC SMAGORINSKI MODEL EDDY VISCOSITY SGS ETC.
In this investigation, large eddy simulations are used to study the dynamics of large and very-large scale motions. 
The equations for the filtered motions are the usual ones \cite[see e.g.][]{Deardorff1970,Pope2000}:
\begin{equation}
\label{eq:LES}
 \frac{\partial \overline{u}_{i}}{\partial x_i} = 0;~~~~~ 
 \frac{\partial  \overline{u}_{i}}{\partial t} + \overline{u}_{j} \frac{\partial \overline{u}{_i}}{\partial x_{j}} = -\frac{\partial \overline{q}} {\partial x_{i}} + \nu \frac{\partial^2 \overline{u}_{i}} {\partial x^2_{j}}-\frac{\partial \overline{\tau}^r_{ij}}{\partial x_{j}},
 \end {equation}
where the overhead bar denotes the filtering action, 
$\overline{\mathbf \tau}^r = \overline{\mathbf \tau}^R - tr(\overline{\mathbf \tau}^R)\, \mathbf{I} /3$,
with
$\overline{\tau}^R_{ij}=\overline{u_i u_j}-\overline{u}_i \overline{u}_j$
and $\overline{q}=\overline{p}+ tr(\overline{\mathbf \tau}^R) / 3$.
The anisotropic residual stress tensor $\overline{\tau}_{ij}$ is modelled choosing an appropriate subgrid model in terms of eddy viscosity $\nu_t$ as
$\overline{\tau}_{ij}^r =-2\nu_t \overline{S}_{ij}$,
where $\overline{S}_{ij}$ is the rate of strain tensor associated with
the filtered velocity field.
For the eddy viscosity, we choose the {\it static} \cite{Smagorinsky1963} model:
\begin{eqnarray} 
\label{eq:Nu_T}
\nu_t =D (C_s \overline{\Delta})^2\overline{\mathcal{S}},
\end{eqnarray}
where  
$\overline{\mathcal{S}} \equiv (2\overline{S}_{ij}\overline{S}_{ij})^{1/2}$, 
$\overline{\Delta}=(\overline{\Delta}_x\overline{\Delta}_y\overline{\Delta}_z)^{1/3}$ is the average length scale of the filter based on the mean grid spacing and $C_s$ is the Smagorinsky constant.
In the following we will use as a reference value $C_s=0.05$, as
\cite{Hwang2010b,Hwang2011}. This value is known to provide the best turbulence statistics compared to those of direct numerical simulations \citep[i.e. the best performance for {\it a posteriori} tests as discussed by][]{Hartel1998}.
To avoid non-zero residual velocity and shear stress at the wall we use the wall (damping) function 
$D=1-e^{-(y^+/A^+)^2}$ proposed by \cite{Menon1999} with $A^+=25$.

The use of the static Smagorinsky model ensures that the residual motions cannot transfer energy to the filtered motions, i.e. there is no `backscatter' of energy. 
This is essential in our approach of determining if large-scale motions can be sustained in the absence of forcing by smaller-scale motions.  
In particular, in the following we will use the same technique used by \cite{Hwang2010b,Hwang2011} to quench small-scale motions, and investigate if the large-scale motions survive despite this quenching.
The technique is simply based on increasing the Smagorinsky constant $C_s$, which is equivalent to an increase of the `Smagorinsky mixing length' $l_0 = C_s \Delta$, as shown by \cite{Mason1986}.
As $C_s$ is increased, therefore, an increasing range of small-scale motions become inactive as they are modelled with an increasingly large (positive) eddy viscosity. 
In `over-damped' simulations the value of $C_s$ will be increased to values in a typical range $C_s \approx 0.1 -  0.18$. 
Two warnings must be issued here in order to avoid possible misunderstanding of the `over-damped LES' technique.
First, it must be reminded that the scope of using `over-damped LES' is, of course,  not to provide a quantitatively accurate representation of large-scale motions (which can be done using state-of-the-art models for the LES) but to show that large-scale motions can survive in the absence of the near-wall process.
For e.g. $C_s=0.1$ or $C_s=0.14$, the near-wall region is (and must be) inaccurate because the buffer-layer cycle has been shut down. However, as already found by \cite{Hwang2010b,Hwang2011} and \cite{Hwang2015} and in the following,  the over-damped flow displays key features of the `real' large-scale motions despite the inaccuracy of the near-wall region.
Secondly, it must be noted that increasing $C_s$ is not equivalent to a reduction of the `effective' Reynolds number in the simulations because in the Smagorinsky model the viscosity depends on the local rate of strain (while it is constant in the Navier-Stokes equations). In practice it is observed that the friction Reynolds number of over-damped solutions is only weakly affected by the increased $C_s$.

\begin{table}
 \begin{center}
\begin{tabular}{lccccccccc}
 \hline\hline 
\noalign{\smallskip}
Name & $L_{x}$ & $L_{z}$ & $N_{x}$ & $N_{y}$ & $N_{z}$  & $\Delta_x$ & $\Delta_{y,\min}$ & $\Delta_{y,\max}$ & $\Delta_{z}$ \tabularnewline
\hline\hline \noalign{\smallskip}
\VLSMBOX&131.9 h	& 18.85 h	& 640 	& 49 	& 128  	& 0.2 h	& 0.01 h		& 0.08 h	& 0.15 h\\
\LSMBOX,& 10.89 h 	& 5.46 h	& 54 	& 49 	& 36  & 0.2 h  	& 0.01 h	 & 0.08 h	& 0.15 h\\ %& 14.3 &   7.0\\
\LSMBOX-Newton,& 10.89 h	& 5.46 h	& 32 	& 61 	& 32  & 0.34 h  	& 0.0055	 h & 0.06 h	& 0.165 h\\%& 43.2 & 21.0\\
\hline\hline
\end{tabular}
 \caption{Numerical domains and discretization parameters used in the present study.
The \VLSMBOX\, and the \LSMBOX\, are used to perform the numerical experiments respectively described in \refsec{SSPLSM} and appendix~A. 
}
 \label{tab:BOXES}
\end{center}
\end{table}

Large eddy simulations are performed with the code {\tt diablo}  \citep{Bewley2001} which implements the fractional-step method based on a semi-implicit time integration scheme and a mixed finite-difference and Fourier discretization in space.
The computational domain, which extends from $0$ to $L_x$ and from $0$ to $L_z$ in the streamwise and spanwise directions respectively and from $-h$ to $h$ in the wall-normal direction, is discretized with $N_x \times N_y \times N_z$ points in respectively the streamwise ($x$), wall-normal ($y$) and spanwise ($z$) directions.
Grid stretching is applied in the wall-normal direction in order to refine the grid near the wall.
No-slip boundary conditions are applied at the walls and periodic boundary conditions at other boundaries.
The numerical domains and the associated discretization parameters used in the following are listed in \reftab{BOXES}.

%====================================================================
\subsection{Computing three-dimensional nonlinear steady solutions}
\label{sec:back_Newton}

In the second part of this study (\refsec{FTW}), we compute steady three-dimensional nonlinear solutions of the filtered equations by
using a modified version of the code {\tt peanuts} \citep{Herault2011,Riols2013}, which provides an implementation of a modified Newton-Krylov method through the PETSc toolkit \citep{petsc}.
An important feature of the algorithm is that it is matrix-free and therefore only relies on (many) calls to the large-eddy simulation solver {\tt diablo}. 

For a Newton-iteration based method to converge, it is crucial to have a good initial guess for the solution. 
One possibility to  obtain such an initial guess is to compute the edge state of the system, i.e. the relative attractor on the surface which (in phase
space) is the boundary of the basin of attraction of self-sustained motions. In cases where the dimension of this boundary is $N-1$, $N$ being
the dimension of the phase space, it is possible to constrain the solution to remain in a  neighbourhood of the boundary by a one parameter bisection technique \citep[see e.g.][]{Itano2001,Toh2003}, sometimes labelled as `edge tracking'.
Edge-tracking has indeed been successfully used by e.g. \cite{Viswanath2007} and \cite{Schneider2008} to compute lower-branch non-trivial steady solutions of the Navier-Stokes equations (unfiltered motions) in plane Couette flow.
In the present study, edge-tracking is implemented by selecting as initial condition of the LES the mean flow profile $\basevel$ to which we add a coherent, three-dimensional large-scale perturbation $\vel_0$ of amplitude $A_0$:
$\vel_0=\basevel + A_0 \vel_0$.
A standard bisection is then performed to compute the threshold value of $A_0$ which separates solutions evolving to active large-scale motions from the laminar solution, as will be further shown in \refsec{FTW}.

%%%%%%%%%%%%%%%%%%%%%%%%%%%%%%%%%%%%%%%%%%%%%%%%%%%%%%%%%%%%%%%%%%%%%%%%%%%%%%%
%%%%%%%%%%%%%%%%%%%%%%%%%%%%%%%%%%%%%%%%%%%%%%%%%%%%%%%%%%%%%%%%%%%%%%%%%%%%%%%
\section{Self-sustained nature of large scale motions} 
\label{sec:SSPLSM}

In the first part of this study we address the question of the self-sustained (or not) nature of motions at large scale in the fully developed turbulent Couette flow at $\Rey=2150$, corresponding to $\Retau=128$.
This Reynolds number, though not excessively high, is large enough to distinguish the size of near-wall structures from those of large-scale motions.

%=================================================================================================
\subsection{Reference large eddy simulations}

As a first step, we assess the ability of our large eddy simulations to reproduce the main features of near-wall, large and very-large scale motions in turbulent Couette flow by performing a reference simulation with the Smagorinsky constant fixed to its reference value $C_s=0.05$.
The simulations are performed in a very large domain  $L_x\times L_z=132\,h \times 18.85\,h$ (labelled \VLSMBOX\, in \reftab{BOXES}), the size of which is comparable to the largest ones used  by \cite{Tsukahara2006} in their direct numerical simulations of turbulent Couette flow. 
Given the typical streamwise and spanwise size of the very-large scale structures \cite[$\lambda_x\simeq 42h - 64h$ and $\lambda_z\simeq 4 h - 5 h$, according to][]{Tsukahara2006}, the domain allows for two or three very-large scale  structures in the streamwise direction and three or four in the spanwise direction. 
The grid spacings in the streamwise and the spanwise direction are set to $\Delta x^+=26.3$ and $\Delta z^+=18.2$, respectively (before dealiasing), while the wall-normal grid spacing is chosen such that the minimum and maximum spacings respectively become $\Delta y^+_{min}=1.27$ and $\Delta y^+_{max}=9.7$. 
We note that while grid spacing is small enough to reasonably resolve the near-wall structures \cite[]{Hartel1998}, 
it is much coarser than the one use by \cite{Tsukahara2006} (who use $\Delta x^+=8$ and $\Delta z^+=6$ and $\Delta y^+= 0.2-5.7$).

\begin{figure}
 \centering
 \includegraphics[width=0.49\columnwidth]{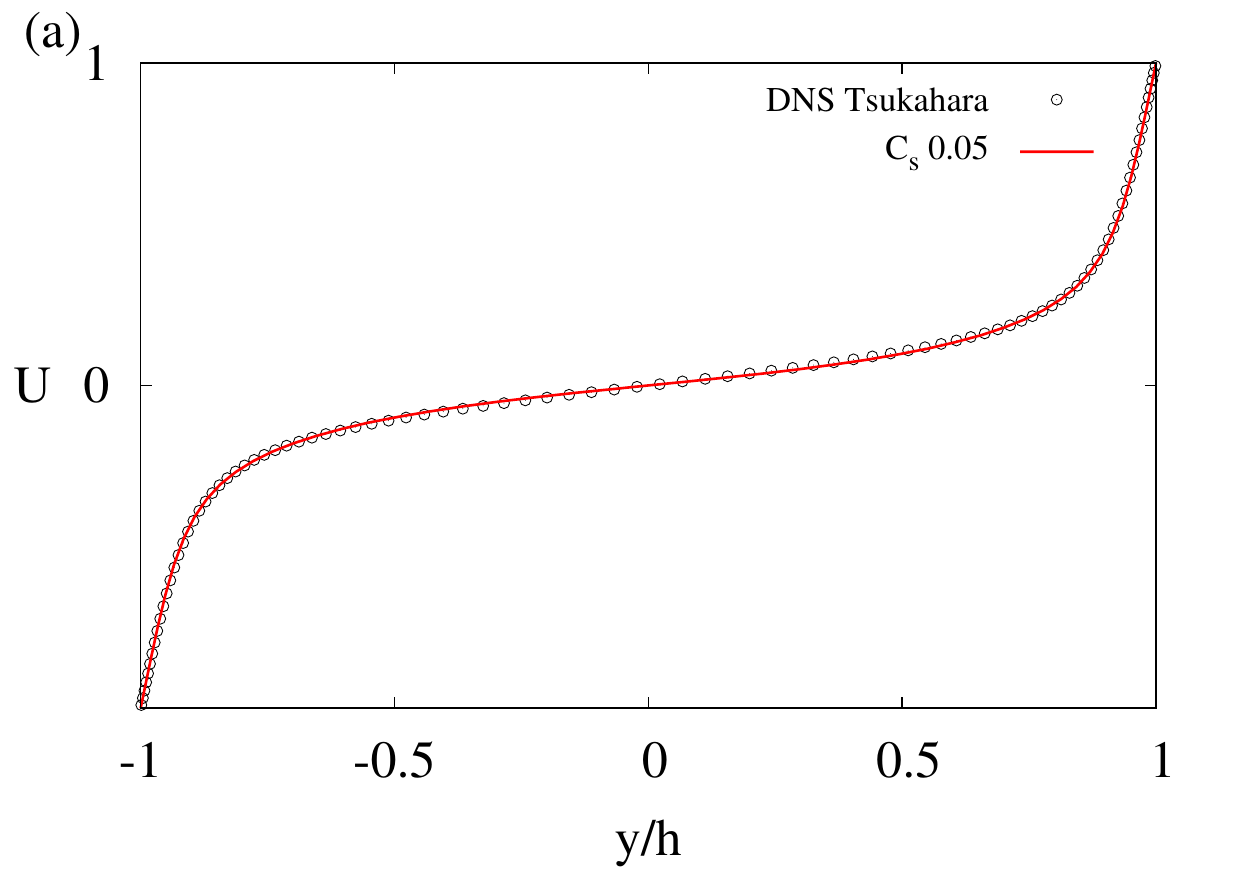}
 \includegraphics[width=0.49\columnwidth]{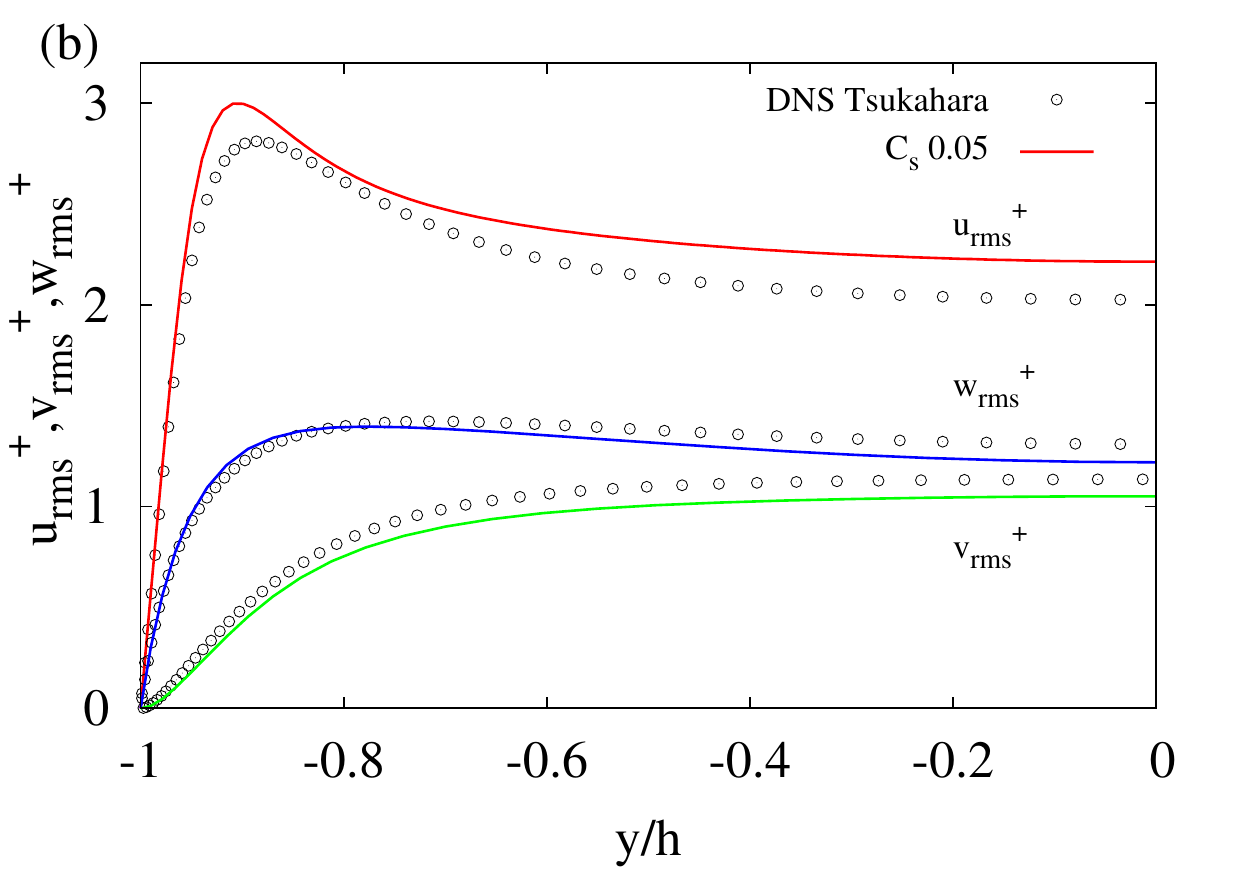}
 \caption{Comparison of the reference LES ($C_s=0.05$, lines) results obtained in the \VLSMBOX\, to the data (symbols) of \cite{Tsukahara2006} obtained by DNS in domains of similar extension at the same Reynolds number  Re=2150 ($Re_{\tau} =128$). 
The mean flow profiles $U^+$ are compared in panel $a$  while in panel $b$ are reported the $rms$ velocity profiles $u'^+_{rms}$, $v'^+_{rms}$  and $w'^+_{rms}$.
}
 \label{fig:DNSLES}
\end{figure}

\begin{figure}
\centering
%\vspace{-7mm}
\includegraphics[width=0.75\textwidth]{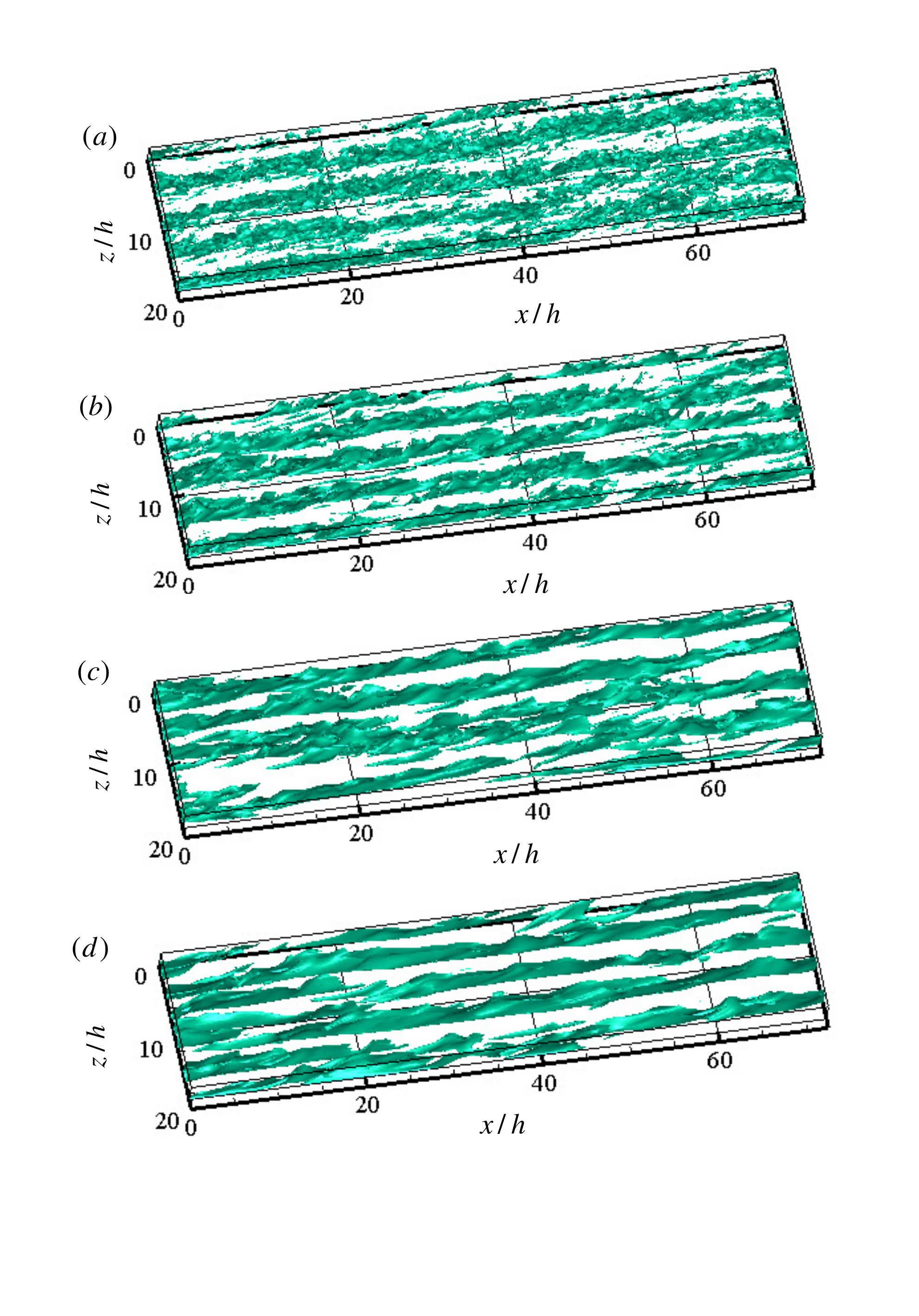}
%\vspace{-25mm}
\caption{
Iso-surface ($u^+=-2.5$) of the instantaneous negative streamwise velocity
fluctuations (low speed streaks) on the lower wall at $\Rey=2150$: (\textit{a}) $C_s=0.05$ (reference
case); (\textit{b}) $C_s=0.10$;  (\textit{c}) $C_s=0.14$; (\textit{d})
$C_s=0.18$.
The visualization illustrates that the progressive quenching of
small-scale motions for increasing values of $C_s$ does not affect the
self-sustainment of large and very large-scale streaky motions.}
\label{fig:Streaks}
\end{figure}

The mean-velocity and $rms$ velocity profiles of the reference LES are in reasonably good agreement with the DNS data of \cite{Tsukahara2006}, as shown in  \reffig{DNSLES}.
In the instantaneous field visualised in \reffig{Streaks}$(a)$, very
long streaky structures with the spanwise spacing roughly at $\lz
\approx 4 h -5 h$ are clearly visible throughout the entire domain
on top of small-scale background turbulence, in agreement with previous simulations \cite[see e.g.][]{Komminaho1996,Tsukahara2006}.
The large-scale streaks also show the characteristic `ramp' structure of the low-speed regions with angles $\approx 10^o - 12^°$ to the wall (see \reffig{sideview}) already observed e.g. in turbulent boundary layers \cite[see e.g. figure~8 of][]{Dennis2011b}.

One-dimensional spectra of the streamwise velocity shown in \reffig{PremSpec}
are also in good agreement with the DNS ones reported by \cite{Tsukahara2006}. 
The spanwise spectra reveal two well-separated peaks depending on the wall-normal location (\reffig{PremSpec}$a$).
The first peak appears at $\lambda_z^+\simeq 100$ in the near-wall region ($y^+\leq 30$), while the second peak is at $\lambda_z \approx 4  h -5 h$ in the outer region ($y \geq 0.2h$);
the former corresponds to  near-wall streaks whereas the latter corresponds to large and very-large scale streaky structures.
The streamwise spectra have a peak at $\lambda_x^+\simeq 700$ in the near-wall region, which is representative of near-wall streaks.
In the outer region, the spectra exhibit two peaks, one  at
$\lambda_x \approx 10h$ and the other at $\lambda_x \approx 42h-64h$. 
The two peaks at $\lambda_x \approx 10h$ and $\lambda_x \approx
42h - 64h$, which are also in good agreement with the DNS results,
correspond respectively to large (LSM) and very large-scale motions (VLSM).

A terminological note is necessary here.
We choose to use the LSM and VLSM terminology also in Couette flow, where it does not appear to be currently standard, because we think that the mechanisms underlying their dynamics are common to other wall-bounded turbulent shear flows.
Also, we will denote by VLSM only structures that have a very large  (e.g. $\approx 40-60h$)  but {\it finite} streamwise wavelength. 
These motions can be detected only in very long domains.
Structures with infinite wavelength in the streamwise direction will not be referred to as VLSM because 
streamwise uniform motions can not self-sustain \citep[see e.g.][]{Waleffe1995} and therefore are not of interest here.

%==============================================================================
\subsection{Overdamped simulations: self-sustained large-scale motions}

Having validated the ability of the reference LES to quantitatively reproduce the main features of the turbulent Couette flow obtained by DNS in previous studies, we now examine if the motions at large scales persist even in the absence of smaller-scale active motions in the near-wall region. 
As in \cite{Hwang2010c}, we therefore gradually increase the Smagorinsky constant $C_s$, which determines the filter width of the LES, to remove the small-scale active structures from the near-wall region.
It should be noted that the increase of $C_s$ does not significantly affect the friction Reynolds number $\Retau=128$ of the reference simulation ($C_s=0.05$): it is indeed found that
$\Retau=119$ for $C_s=0.10$, $\Retau=117$ for $C_s=0.14$, $\Retau=119$ for $C_s=0.18$. 

\begin{figure}
\centering
\includegraphics[width=0.99\textwidth]{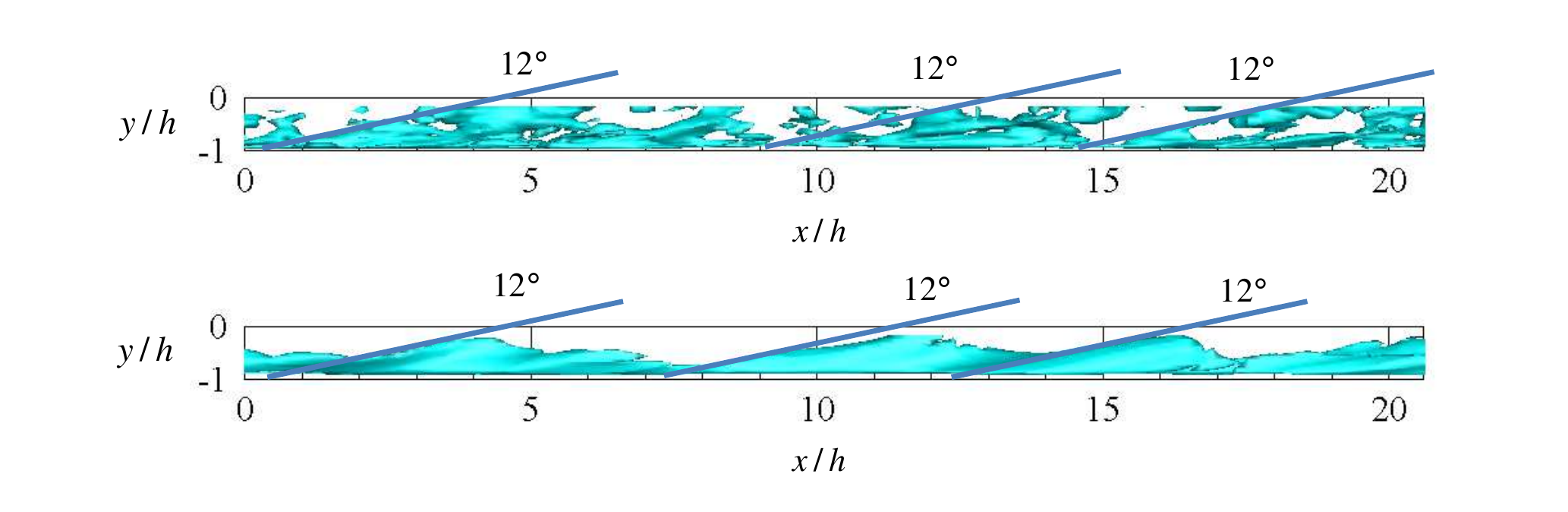} 
%\vspace{-5mm}
\caption{Side view of the low velocity fluctuations in the \VLSMBOX\, for the reference case $C_s=0.05$ (top panel) and the over-damped $C_s=0.14$ (bottom panel) at $\Rey=2150$.
The data are the same as in panels $a$ and $c$ of \reffig{Streaks}. 
In order to highlight the ramp structure of large-scale motions, only a limited streamwise part of the computational domain is represented and the levels $u^+=-3.9$ for $C_s=0.05$ and $u^+=-5.3$ for $C_s=0.14$ are used.
The $12^o$ angle is measured from the wall. }
\label{fig:sideview}
\end{figure}

The flow fields visualised in \reffig{Streaks} show that the small-scale turbulent motions are gradually damped out as $C_s$ is increased. 
For $C_s=0.14$ (\reffig{Streaks}$d$), only the motions at large and very large scale are left in the flow field, suggesting the outer motions are likely to be sustained by themselves. 
The survival of the large-scale motions persist at $C_s=0.18$ (see \reffig{Streaks}$d$), although they are eventually quenched for $C_s=0.25$ (not shown).
The examination of \reffig{sideview} further confirms that when small-scale motions are smoothed out the surviving coherent large and very-large scale structures are reasonably similar to the `natural' ones obtained in the reference case.
In particular, they show the typical `ramp' structure of the low-speed regions with angles $\approx 10^o - 12^o$ to the wall.
\begin{figure}
%\vspace*{2mm}
\centering
  \includegraphics[width=0.8\columnwidth]{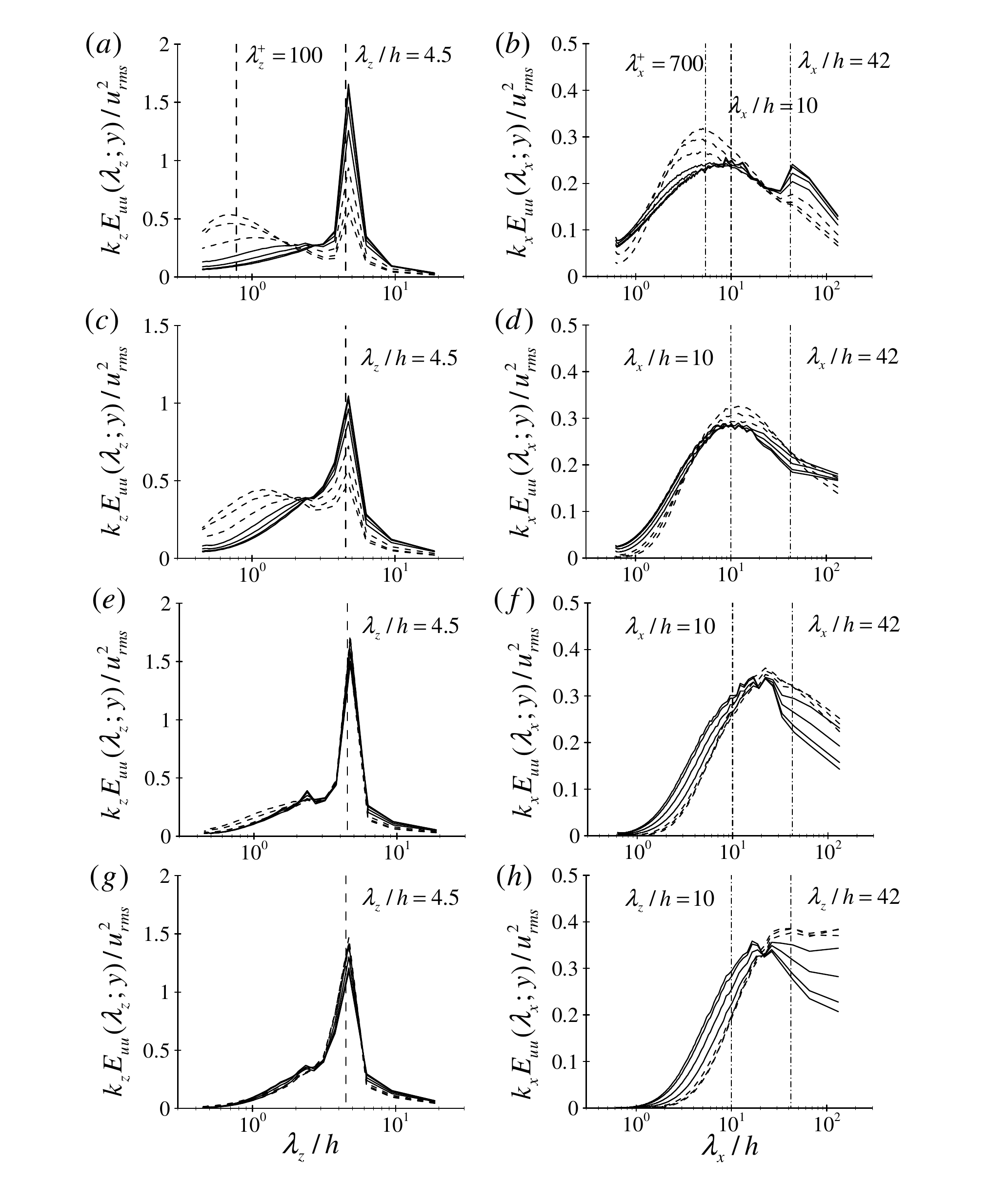}
\caption{    Spanwise premultiplied power spectra   $k_z E_{uu} (\lz)$ [(a), (c), (e), (g)] and
   streamwise premultiplied power spectra $k_x E_{uu} (\lx)$ (panels $b$, $d$, $f$, and $h$) for respectively the reference simulation with  $C_s=0.05$ (panels $a$ and $b$)  and for the cases
   $C_s=0.10 $ ($c$ and $d$) and
   $C_s=0.14 $ ($e$ and $f$) and
   $C_s=0.18 $ ($g$ and $h$).
   The premultiplied spectra are extracted in the inner- ($y^+=11, 17, 28$ based on $Re_{\tau}$ of the reference simulation, dashed lines)
   and in the outer layer (at distances from the bottom wall of $y+h=0.40 h, 0.57 h, 0.78 h, h$ solid lines).}
\label{fig:PremSpec}
\end{figure}

The premultiplied spectra reported in \reffig{PremSpec} provide a quantitative confirmation of this scenario. 
The original ($C_s=0.05$) peak location of the near-wall motions at $\lambda_z^+\simeq 100$ 
 in the spanwise spectra (\reffig{PremSpec}$a$) is initially shifted to a larger spanwise wavelength for $C_s=0.10$  (\reffig{PremSpec}$c$) and completely suppressed for $C_s=0.14$  (\reffig{PremSpec}$e$) and higher (\reffig{PremSpec}$g$), where only the footprint of large-scale motions appears in the near wall region. 
Similar features are observed in the streamwise spectra where the original peak location at $\lambda_x^+\simeq 700$ in the near-wall region (\reffig{PremSpec}$b$) is shifted to higher wavelengths up to its complete suppression  (\reffig{PremSpec}$d$, $f$ and $h$). 

In contrast, large-scale motions (LSM) appear to be fairly robust to the increase of $C_s$. 
The peak at $\lambda_z\simeq 4.5h$ in the spanwise spectra robustly remains at the same location for all increasing values of $C_s$ (\reffig{PremSpec}$a$, $c$, $d$ and $e$). 
In the streamwise spectra, for $C_s=0.10$ the LSM peak at $\lambda_x\simeq 10h$ remains unchanged, while the VLSM peak at $\lambda_x \simeq 42h - 64h$ is blurred.  
For $C_s=0.14$, the original location of the first peak ($\lambda_x\simeq 10h$) has drifted to a slightly longer streamwise wavelength ($\lambda_x\simeq 14 h$) (\reffig{PremSpec}$f$), and it appears that this feature makes the presence of the VLSM structures even more blurred. 
Similar drifts of the streamwise wavelength of LSM for relatively large $C_s$ have also been observed in the case of the pressure-driven channel flow \citep[][]{Hwang2010c}. 
When $C_s=0.18$, the artificial over-damping begins to have a non-negligible effect on the dynamics of large-scale motions in the near-wall region where energy accumulates at the longest available streamwise wavelengths. 
This is why in \reffig{sideview} and in the following we only discuss the dynamics of structures corresponding to $C_s \lesssim 0.14$.

These results suggest that motions with typical spanwise spacings $\lambda_z\simeq 4 h  - 5 h$  sustain by a process which is not forced by motions at smaller scales. 
The relation between large-scale (LSM) and very-large scale motions (VLSM) is left partially unsolved by these simulations because they do not exclude that the self-sustained process is associated to the VSLM very long wavelengths. 
The reference and overdamped simulations have been therefore repeated in a smaller domain (the \LSMBOX\, listed in \reftab{BOXES}) having the typical dimensions of large-scale motions. 
The results of this additional series of LES, discussed in appendix~\ref{sec:LESLSMBOX}, are very similar to the ones discussed above and show that large-scale motions self-sustain also in the absence of potentially active motions of even larger scale when near-wall active processes are quenched.
There is therefore an active process at precisely this scale ($\lambda_z \approx 4 h - 5 h$, $\lambda_x \approx 10 h -12 h$). 
The nature of this process is further explored in the next section.

%%%%%%%%%%%%%%%%%%%%%%%%%%%%%%%%%%%%%%%%%%%%%%%%%%%%%%%%%%%%%%%%%%%%%%%%%%%%%%%
%%%%%%%%%%%%%%%%%%%%%%%%%%%%%%%%%%%%%%%%%%%%%%%%%%%%%%%%%%%%%%%%%%%%%%%%%%%%%%%
%\clearpage
\section{Steady 'exact' nonlinear large-scale solutions of the filtered equations}
\label{sec:FTW}

The overdamped simulations of filtered motions have shown that large-scale motions can self-sustain even when near-wall motions are artificially quenched. 
Similarly to what has been done to understand transitional structures for the Navier-Stokes equations, we investigate the specific mechanism by which turbulent large-scale motions sustain by looking for the existence of nonlinear `exact' solutions. 
More specifically,  we look for steady solutions of the filtered equations at Reynolds numbers where the flow is in a fully developed turbulent state. 
To make the steady state computations manageable and to avoid potentially active very-large scale motions to come into the picture we consider
the domain \LSMBOX-Newton listed in \reftab{BOXES}, whose size  $L_x \times L_z =  10.9 h \times 5.5 h$ is that of large-scale motions (see \reffig{PremSpec}).
The dimensions of the \LSMBOX~ also correspond to those of the 'optimum' domain considered by \cite{Waleffe2003} and are very similar to those of domains in which early `transitional' steady solutions of the (unfiltered) Navier-Stokes equations have been found \citep{Nagata1990,Clever1992}.  
We will show that this is not a mere coincidence, and that these solutions, that we will denote by NBCW (Nagata-Busse-Clever-Waleffe), are indeed much more related to self-sustained large-scale motions than to near-wall cycles.

%=================================================================================================
\subsection{Coherent steady solutions at \Rey=750}

We begin by looking for exact solutions near the relatively low Reynolds number $\Rey=750$ (corresponding to $\Retau=52$), which roughly corresponds to twice the value of the transitional Reynolds number.
The box discretization details are reported in \reftab{BOXES}.
It has been verified \citep{Rawat2014b} that reference large eddy simulations at the considered $\Rey=750$ with $C_s = 0.05$ in the \LSMBOX~compare well to the DNS results obtained in the same box and are relatively similar to those obtained in larger domains at the same Reynolds number \cite[see also][for a discussion of the box size effects]{Komminaho1996,Tsukahara2006}. 
The premultiplied spectra also show a reasonable agreement with the DNS results for $C_s=0.05$ and only large-scale motions survive when the near-wall cycle is artificially quenched by increasing the Smagorinsky constant $C_s$ \citep{Rawat2014b}. 

\begin{figure}
 \centering
\includegraphics[width=0.49\columnwidth]{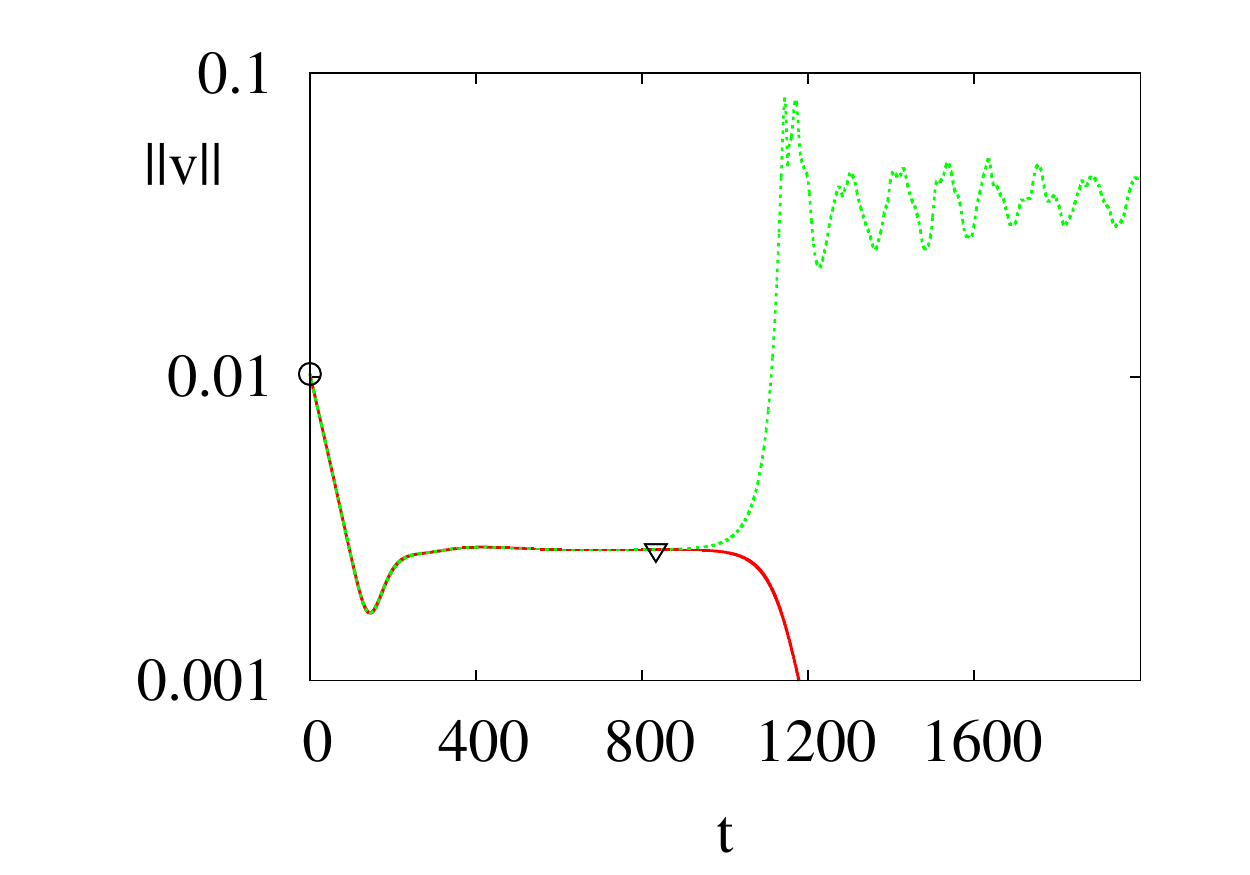} 
 \includegraphics[width=0.49\columnwidth]{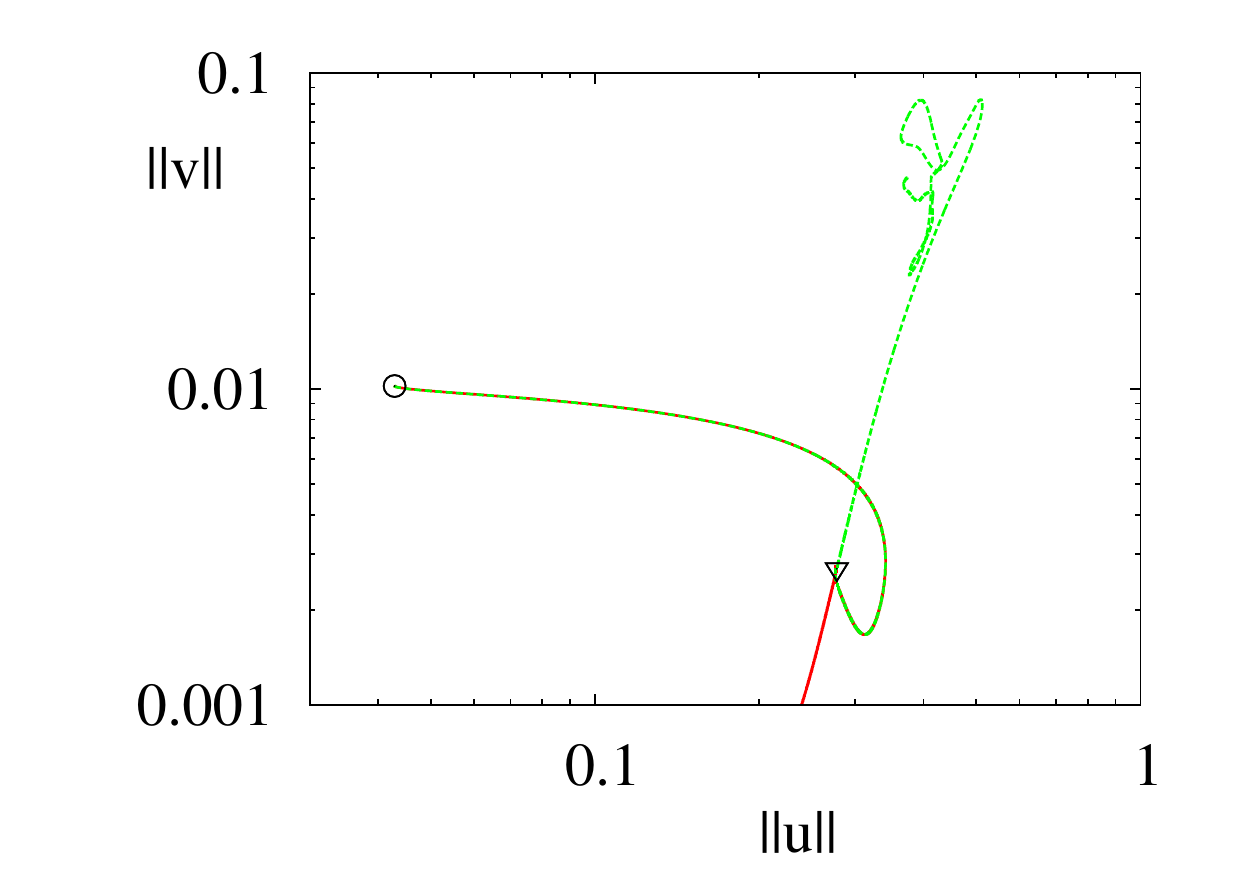} 
 \caption{Edge tracking in the over-damped LES at \Rey=750 ($\Retau =52$) and $C_s=0.14$. 
Two initial conditions (circle symbol) lying almost on the edge of chaos surface initially remain near the surface while being attracted to the steady solution edge state (triangle symbol). 
The solution lying on the upper side of the edge of chaos (green line, dotted) finally evolves toward a state with irregular large-scale active motions while the other one  (solid, red line) finally relaxes to a state with no large-scale active motions.
The trajectories are shown in both the $t - \|v\|$ (top panel) and in the $\|u\| -\|v\|$ plane (bottom panel) displaying the streaks-vortices dynamics. 
} \label{fig:BisectionCs14} \end{figure}

\begin{figure}
 \centering
 \includegraphics[width=0.8\columnwidth]{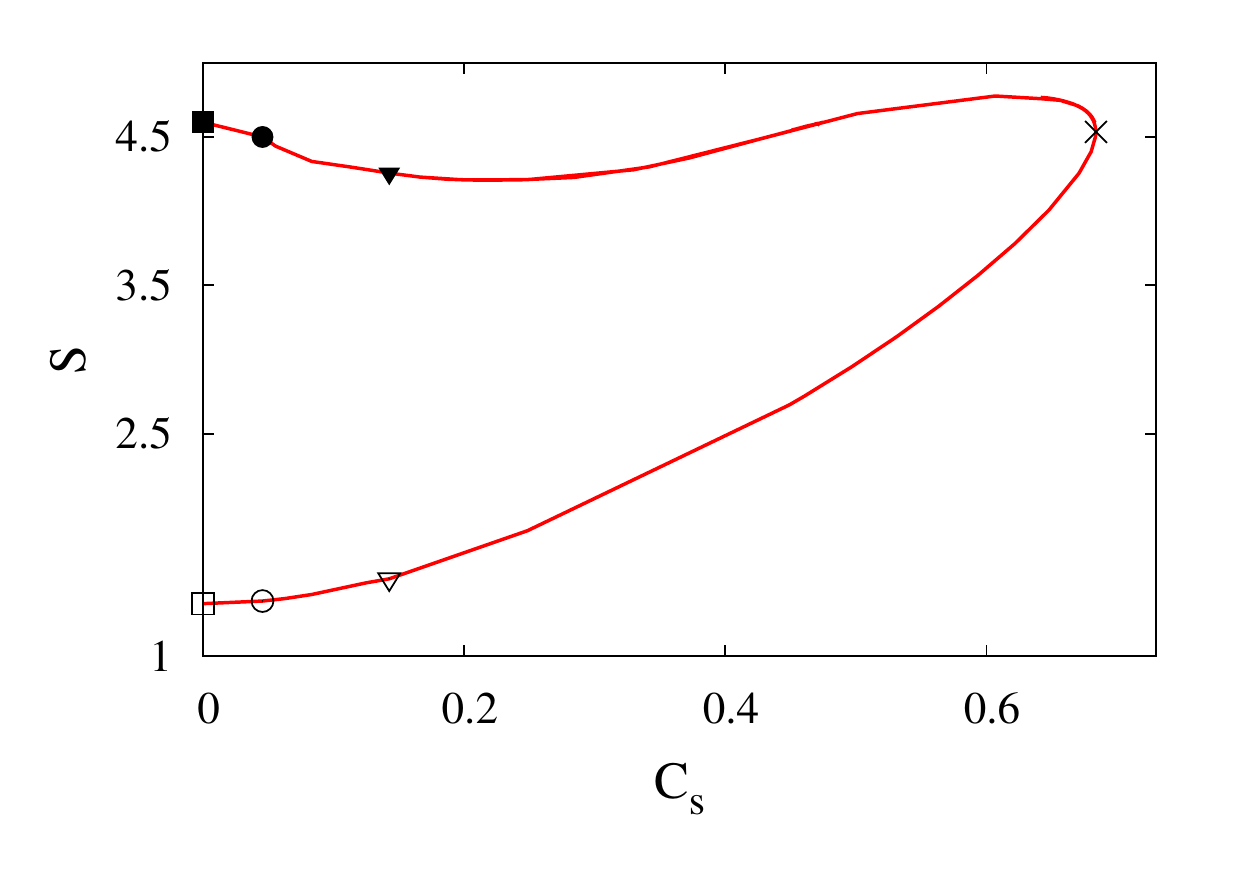}
%\vspace{-5mm}
 \caption{Continuation diagram of the steady solutions in the Smagorinsky constant $C_s$ at $\Rey=750$ ($\Retau=52$). $S=\tau_w / (\mu U_w /h)$ is the wall shear rate of the solution normalized by its laminar value.
The symbols denote the solutions found at $C_s=0$ (Navier-Stokes, squares, with $S=1.37$ on the lower branch and $                   S=4.59$ on the upper branch), $C_s=0.05$ (reference LES, circles) and $C_s=0.14$ (overdamped LES where the edge-tracking has been performed, triangles).
Empty symbols denote lower branch solutions, filled symbols denote upper branch solutions.
}
 \label{fig:CsCONT}
\end{figure}
\begin{figure}
 \centering
 \includegraphics[width=0.32\columnwidth,height=0.32\columnwidth]{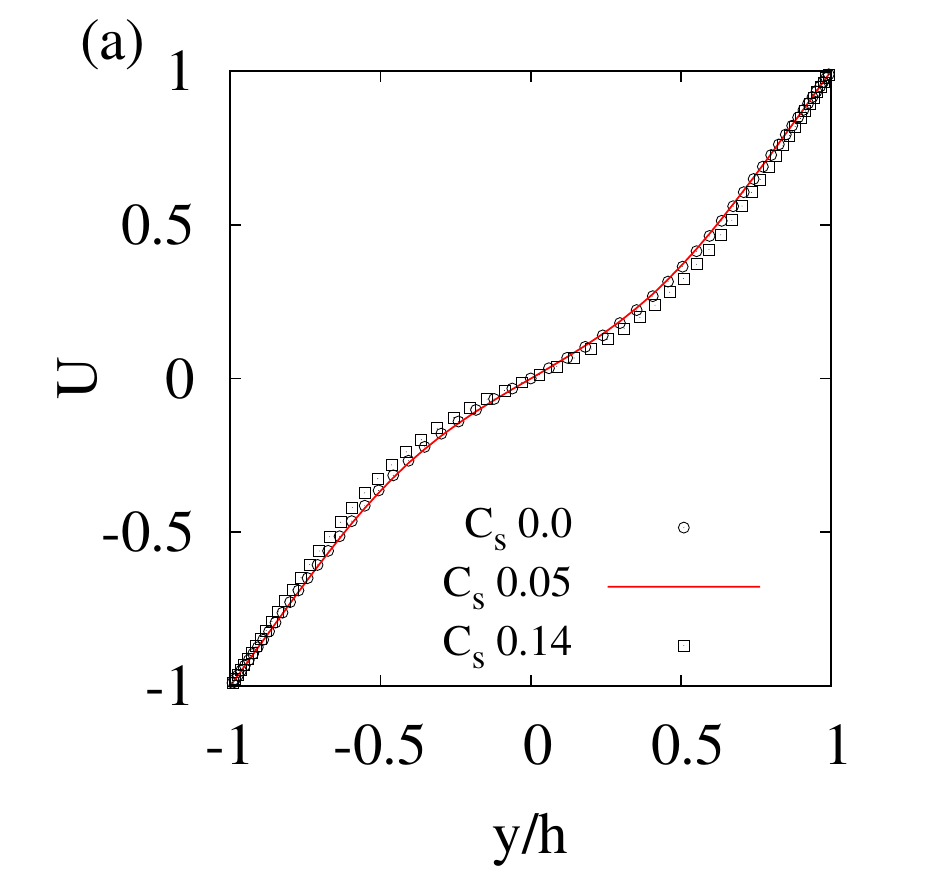}
 \includegraphics[width=0.32\columnwidth,height=0.32\columnwidth]{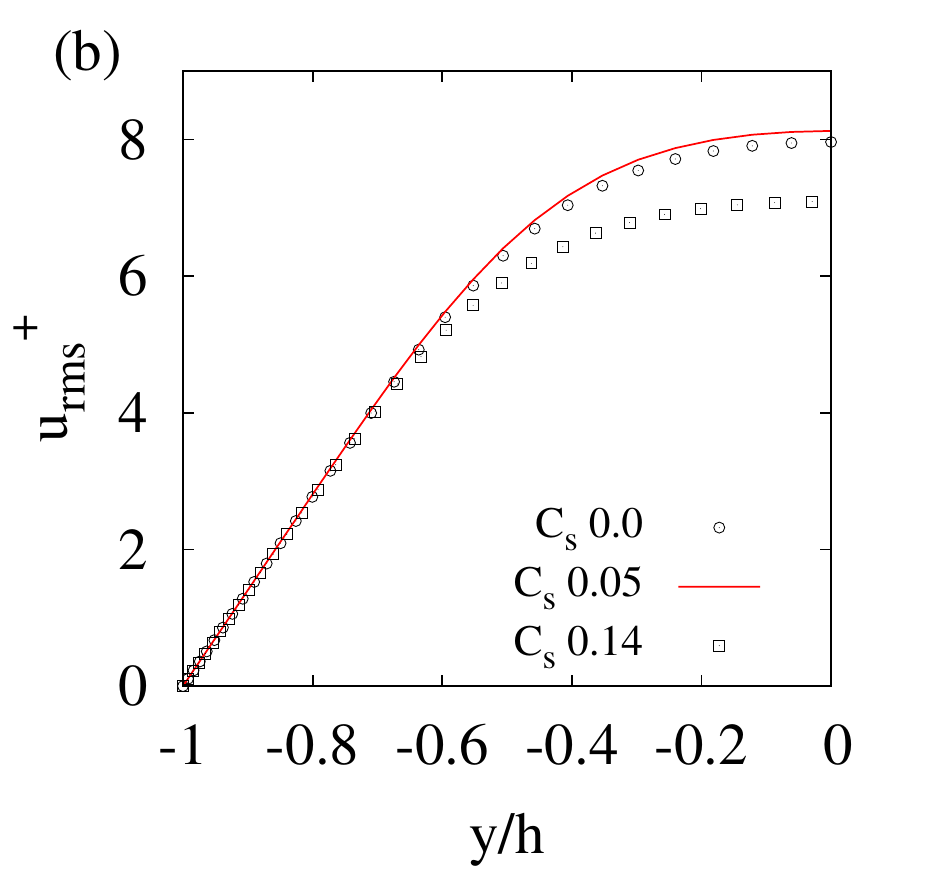}
\hspace{-1mm} 
\includegraphics[width=0.32\columnwidth,height=0.32\columnwidth]{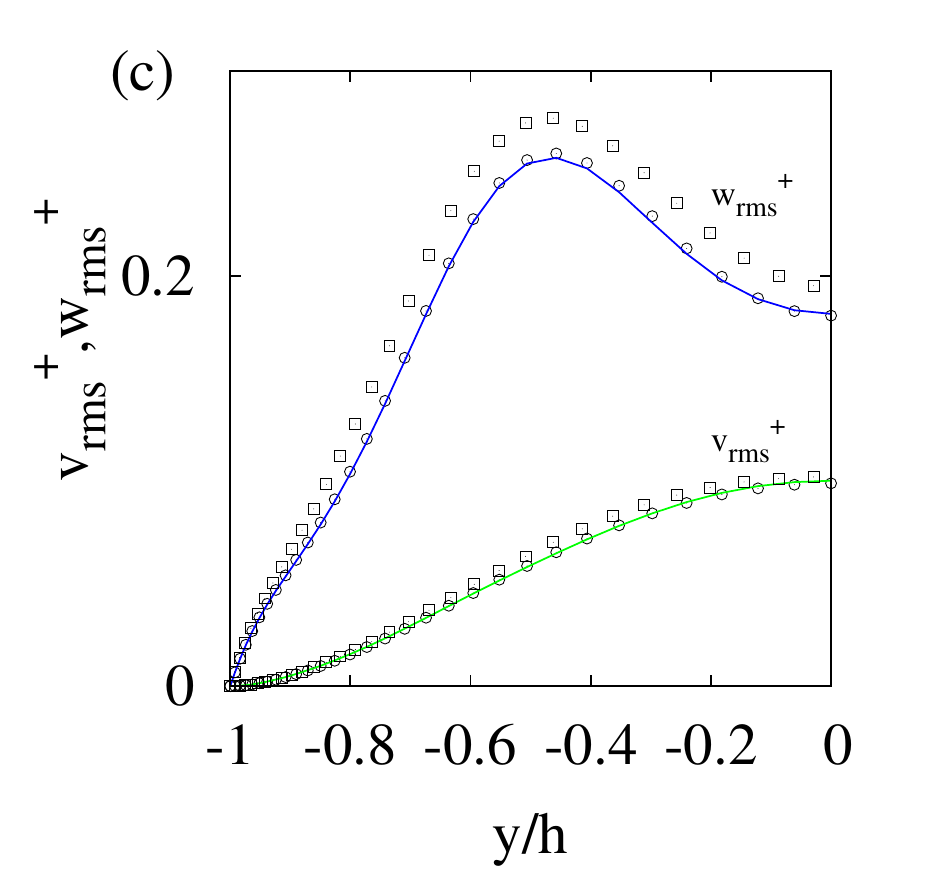}
 \includegraphics[width=0.32\columnwidth,height=0.32\columnwidth]{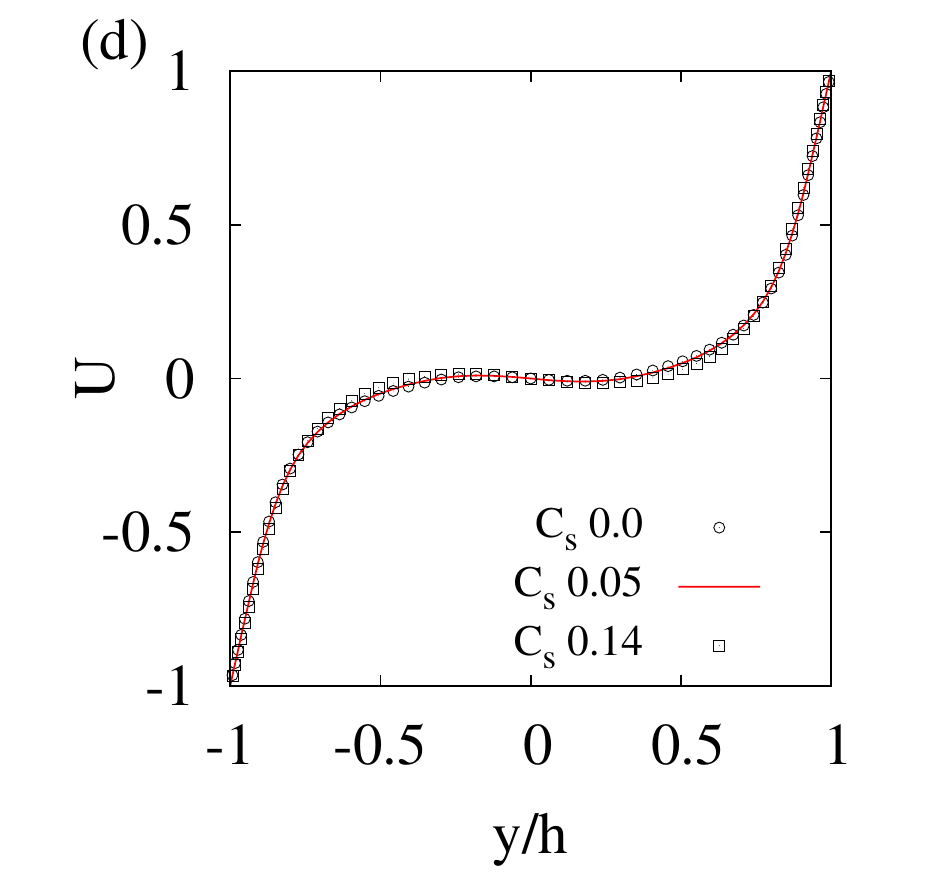}
 \includegraphics[width=0.31\columnwidth,height=0.32\columnwidth]{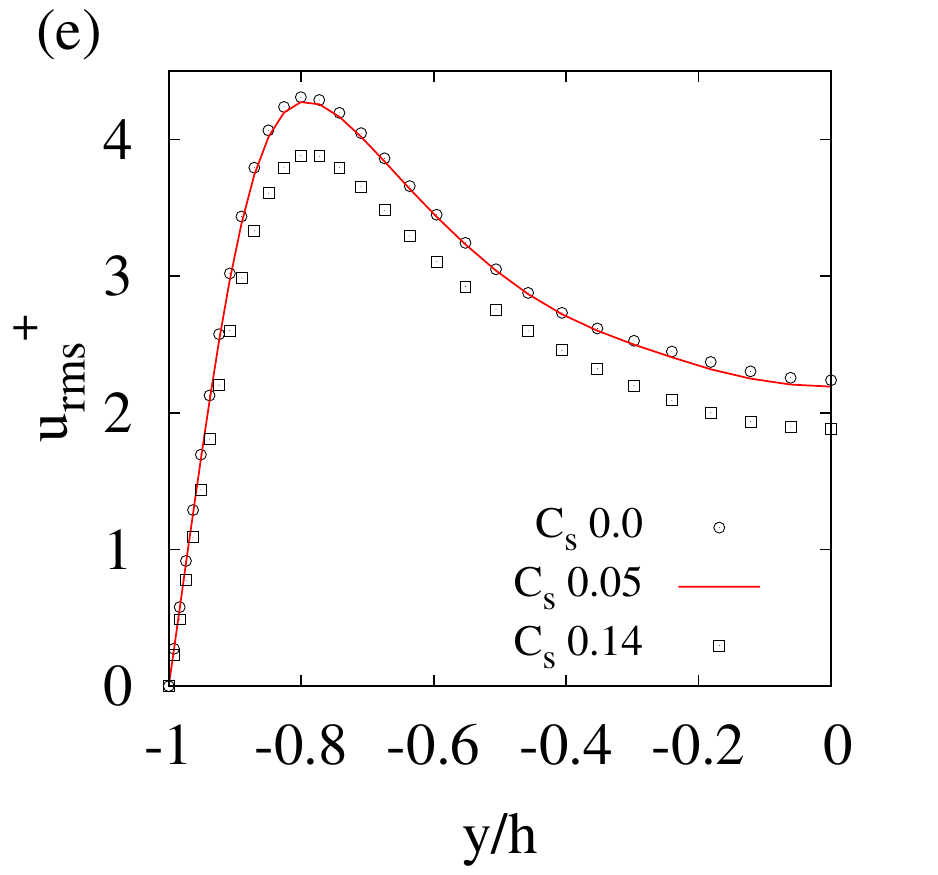}
\hspace{-1mm} 
\includegraphics[width=0.32\columnwidth,height=0.32\columnwidth]{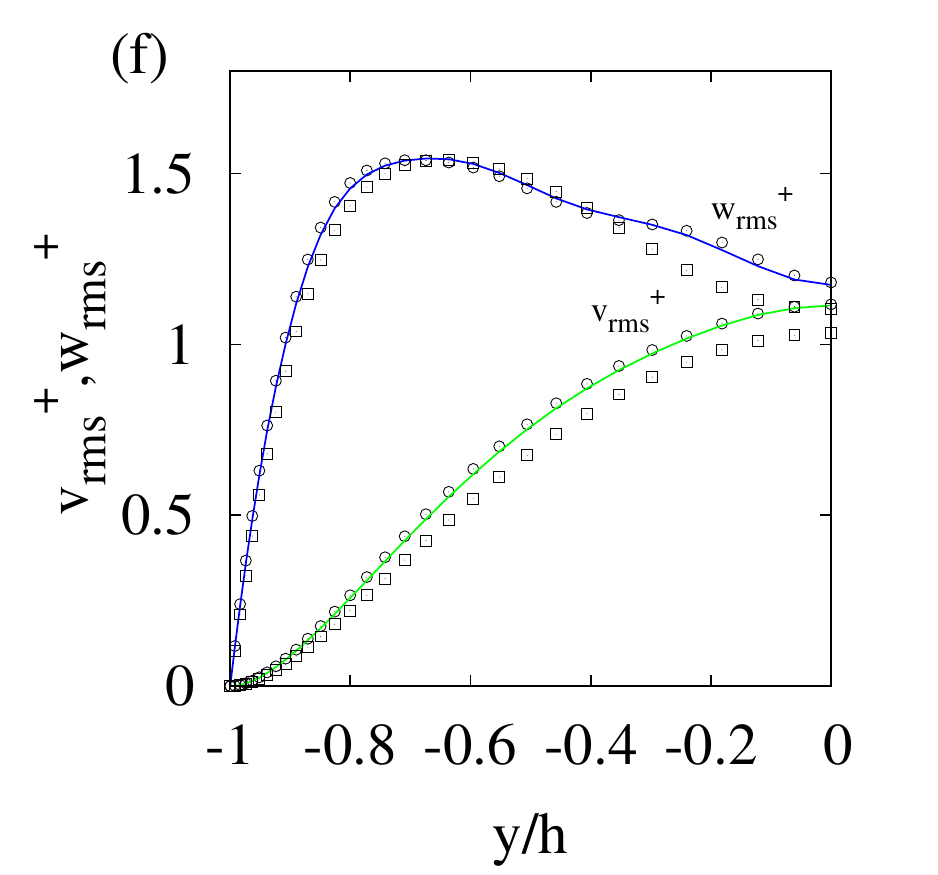}
 \caption{Lower (panels $a$, $b$ and $c$) and upper (panels $d$, $e$ and $f$) branch steady solutions in the \LSMBOX~at \Rey=750 ($\Retau=52$):
mean flow (panels $a$ and $d$), streamwise (panels $b$ and $e$) and cross-stream (panels $c$ and $f$) $rms$ velocity profiles corresponding to  $C_s=0.14$ (overdamped large eddy simulation), $C_s=0.05$ (reference large eddy simulation) and $C_s=0$ (Navier-Stokes steady solution). 
}
 \label{fig:FTWRE750}
\end{figure}

As discussed in \refsec{back_Newton}, an edge-tracking technique can be effective to compute lower-branch steady solutions in plane Couette flow.
We therefore apply the edge-tracking technique to the overdamped LES simulations with
$C_s =0.14$ at \Rey=750 in the \LSMBOX~ where the near-wall cycle is quenched and only large-scale motions survive. 
The initial condition given to the LES simulations consists in the mean turbulent flow to which is added a pair of streamwise-uniform counter-rotating rolls along with a sinuous perturbation of the spanwise velocity, which is similar to that used by \cite{Toh2003} and \cite{Cossu2011}:
${\bfu}_0=
\left\{U(y),0,0\right\} 
+ A_0 \left\{0,\partial \psi_0/\partial z,-\partial \psi_0/\partial y\right\}
+0.1\,A_0 \left\{0,0, w_{sin}\right\}$
where 
$\psi_0(y,z)=\left(1-y^2\right)\, \sin \left(2 \pi z/L_z\right)$ and 
$w_{sin}(x,y)=\left(1-y^2\right)\,\sin \left(2 \pi x/L_x\right).$
The bisection is performed by adjusting the amplitude $A_0$ of the perturbations.
The results of the edge tracking, displayed in \reffig{BisectionCs14}, show that the edge state solution has constant energy. 
The initial transient clearly shows the coherent transient growth of the initial condition (slow decrease of the wall-normal velocity perturbations and high increase of the streamwise streaks norm $\|u\|$) leading to the edge state.

The edge-tracking solution is then used as an initial guess for Newton-Krylov iterations, performed with the \peanuts\, code. 
The solver converges to a steady solution (i.e. with zero phase speed, up to the precision of the Newton solver computation) of the (LES) equations for the filtered motions.
The converged solution can be continued in $C_s$ using a Newton-based continuation. 
When continued to higher $C_s$, the solution encounters a saddle-node bifurcation from which the lower branch and an associated upper branch originate, as shown in \reffig{CsCONT}. 
As the solutions obtained for $C_s \gtrsim 0.18$ lose much of their physical relevance because the LSM become themselves over-damped, the high-$C_s$ continuation should be considered only as a method (just as an homotopy) to access upper branch solutions. 
Also, it must be noted that the high-$C_s$ bifurcation is not simply reproducing the low-$\Rey$ bifurcation of Navier-Stokes solutions. Here the stress tensor depends non-linearly on the rate of strain, which models turbulent dissipation. As a consequence, e.g.  the value of shear parameter $S$ at the the saddle-node bifurcation is more than the double of the value of $S$ at the low-Reynolds number saddle-node bifurcation of Navier-Stokes solutions.

Even more interestingly, both upper and lower branch solutions can be continued to lower values of $C_s$ and in particular to $C_s=0.05$, the `reference' value used to match DNS solutions, and even to $C_s=0$ which corresponds to solutions of the (unfiltered) Navier-Stokes equations.
The mean and $rms$ velocity profiles corresponding to $C_s=0.14$, $C_s=0.05$ and $C_s=0$, reported in \reffig{FTWRE750}, are very similar to one another and do remind the NBCW  solutions of the  Navier-Stokes equations.
For the lower-branch solutions, most of the energy is concentrated in the quasi-streamwise streaky motions (mainly $u_{rms}$ component) which are forced by low-amplitude quasi-streamwise vortices (mainly the $v_{rms}$ component) while for the upper-branch solutions, quasi-streamwise streaks and vortices have comparable energy.

\begin{figure}
 \centering
\vspace{3mm}
 \includegraphics[width=0.85\columnwidth]{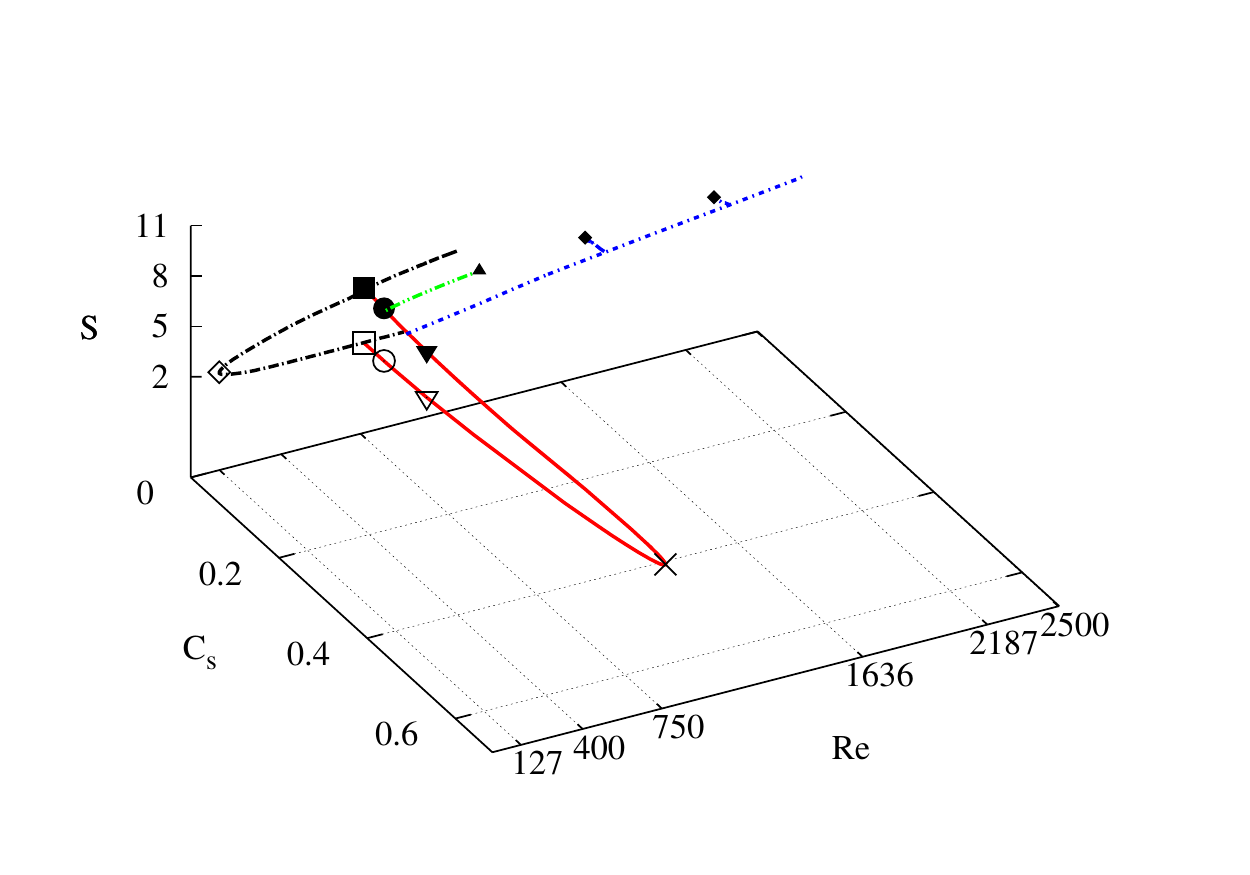}
%\vspace{-5mm}
 \caption{ Continuation diagram of \tws solutions in the $C_s - \Rey$ parameters plane.
The solid (red) line continuation in $C_s$ is that shown in \reffig{CsCONT} (see its caption for the meaning of the symbols).
The continuation in Reynolds number at $C_s=0$ (black line, dashed) shows that the solutions continued from the turbulent case (square symbols) belong to the NBCW   branch of Navier-Stokes  solutions originating in a saddle-node bifurcation at $\Rey =127$ (in this box with $L_x=10.89$ and $L_z=5.46$).
It has been possible to directly continue the $C_s=0.05$ solution from $\Rey=750$ to $\Rey=1164$ (green line connecting to the solution denoted by a filled triangle).
Nonlinear steady solutions at higher Reynolds numbers are found by first continuing 
the solutions obtained for $C_s=0.1$ to higher Re (blue line, dashed-dotted)
and then reducing $C_s$ to  $C_s=0.05$ (filled diamond symbols),
except for the $\Rey=2187$ solution which could only be continued to $C_s=0.06$.
}
 \label{fig:CsReCONT}
\end{figure}
\begin{figure}
 \centering
 \includegraphics[width=0.42\columnwidth,height=0.42\columnwidth]{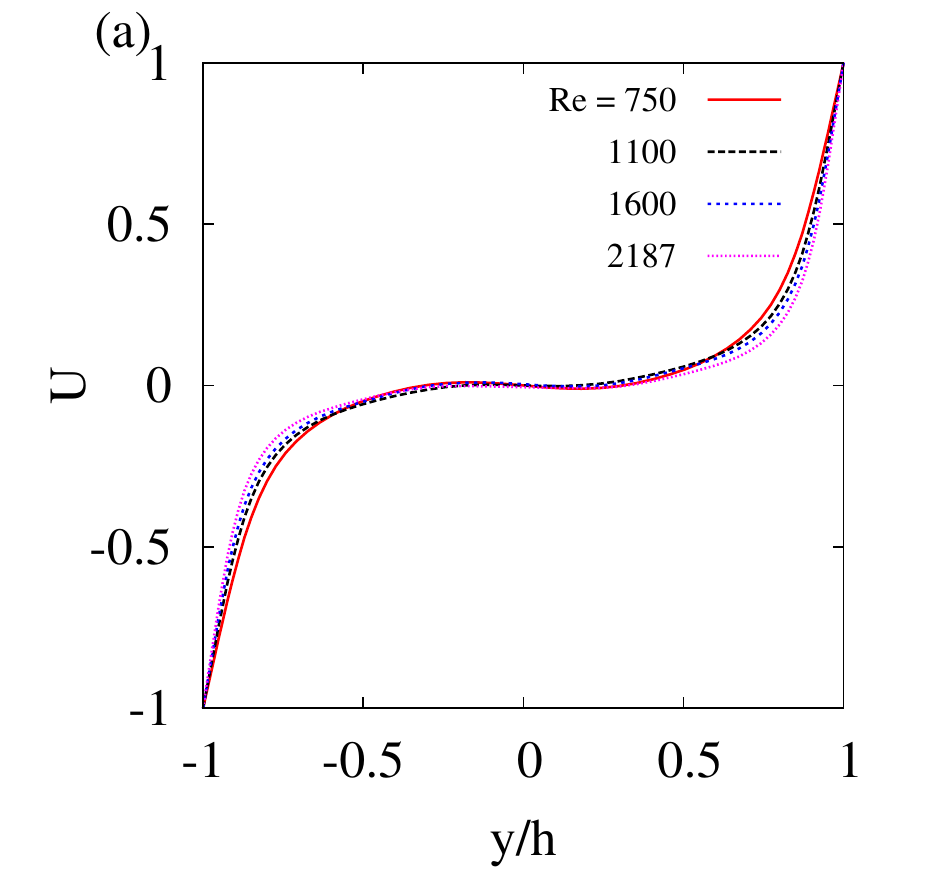}
 \includegraphics[width=0.42\columnwidth,height=0.42\columnwidth]{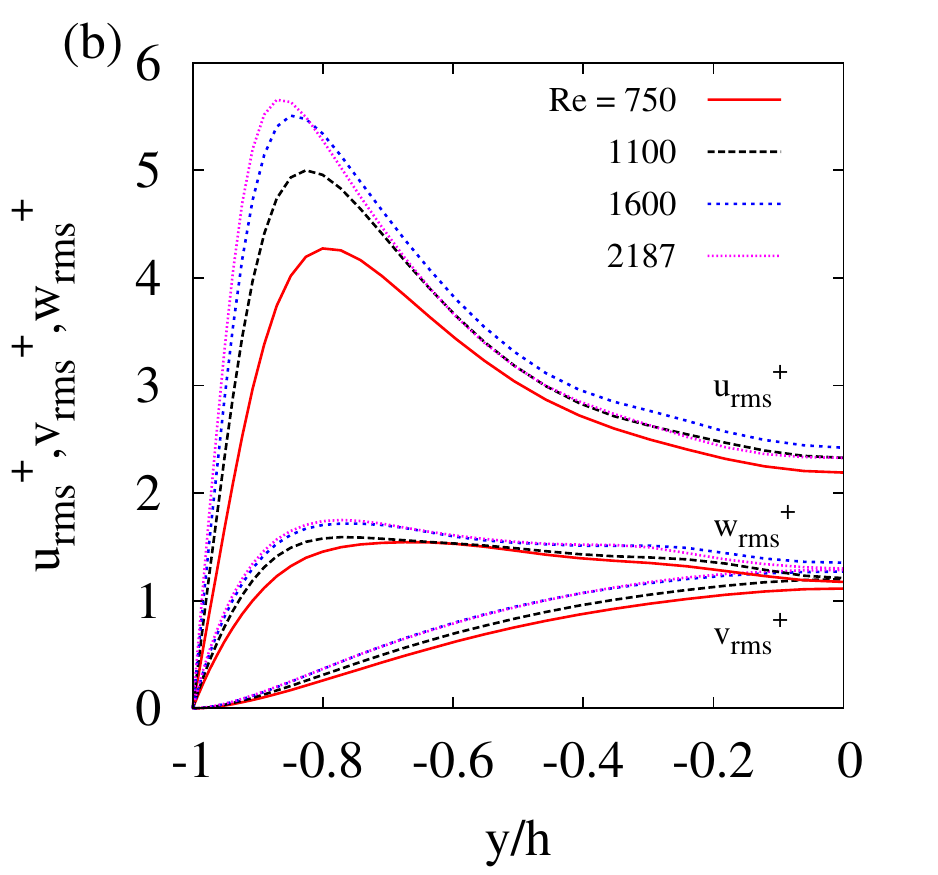}
 \caption{Mean flow (panel $a$) and $rms$ velocity profiles (panel $b$) corresponding to the 
upper branch coherent steady solutions obtained for increasing Reynolds numbers
\LSMBOX\,  ($C_s=0.06$ for $\Rey=2150$ and $C_s=0.05$ otherwise).
}
 \label{fig:HighReNTW}
\end{figure}
\begin{figure}
 \centering
 \includegraphics[width=0.75\columnwidth]{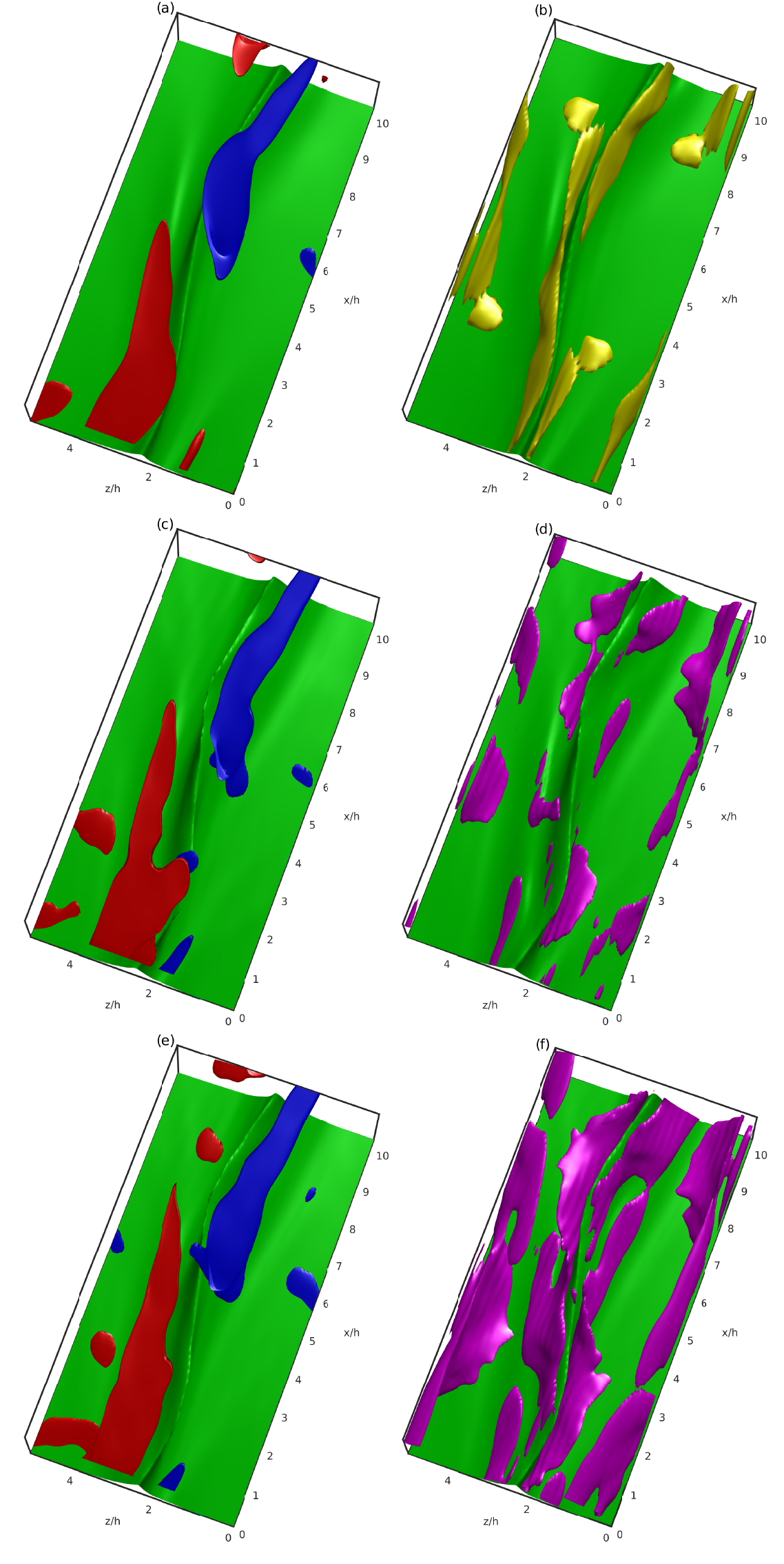}
  \caption{ Visualisation of the upper branch large-scale coherent steady solutions of the LES equations obtained with the reference value $C_s=0.05$ and for increasing values of the Reynolds number: 
\Rey=750 (corresponding to $\Retau=52$, top row),
\Rey=1600 ($\Retau=99$, middle row),
\Rey=2187 ($\Retau=127$, with $C_s=0.06$). 
The panels on the left represent the streaks and quasi-streamwise vortices: the green surface corresponds to a streamwise velocity value $50\%$ of its maximum value while the blue and green surfaces correspond streamwise vorticity values equal to $\pm 70\%$ of the maximum.
The panels on the right represent the streaks and the relative eddy-viscosity associated to the filtered small-scale motions. The green surface is the same as in the left panel, while the violet surfaces correspond to $\nu_t / \nu = 40\%$ and the yellow one to $\nu_t / \nu = 10\%$.}
 \label{fig:UBHighRe}
 \end{figure}

%=====================================================================================================
\subsection{Continuation to higher Reynolds numbers}

The steady solutions obtained at $\Rey=750$ are very similar to the NBCW  solutions. 
We have therefore continued the solution corresponding to $C_s=0$ (i.e. the unfiltered Navier-Stokes solution) from $\Rey=750$ to $\Rey=400$ where it appears to be identical to the one computed by \cite{Waleffe2003} in the same box at the same $\Rey$.
Continuation to lower $\Rey$ indeed shows that the upper and the lower branch are connected by the saddle-node bifurcation at $\Rey=127$ \citep{Waleffe2003} as shown in \reffig{CsReCONT}.  

At $\Rey=750$ ($\Retau \approx 52$), the near-wall and large-scale motions scales are not yet separated enough to allow definitive conclusions on the nature of the steady solutions of the filtered equations.  
It is therefore potentially unclear that the computed steady solutions are associated to the self-sustained large-scale motions discussed in \refsec{SSPLSM} and not to near-wall motions. 
In order to clarify this issue the coherent steady solutions of the over-damped LES have been continued to higher Reynolds numbers, keeping the box dimensions constants in outer units ($L_z=5.5 h$ and $L_x=10.9 h$).

We concentrate on upper branch solutions because they are most relevant to the turbulent dynamics (lower branch solutions are related to the transition problem which is probably less relevant in the turbulent regime).
A continuation to higher $\Rey$ fails to converge for Reynolds numbers larger than $Re \approx 1164$ ($\Retau \approx 74$) for the reference $C_s=0.05$ (but also for $C_s=0$).
To circumvent this difficulty, an alternate path has been followed in the parameter space by first computing solutions at $C_s=0.1$, as shown in \reffig{CsReCONT}. 
The $C_s = 0.1$ steady solutions can be continued up to $Re \approx 2500$. 
At selected Reynolds numbers ranging up to $\Rey=2187$, the $C_s = 0.1$ solutions can then be tracked down to $C_s=0.05$ except for  $\Rey=2187$, corresponding to $\Retau \approx 127$, where the solution can not be continued below $C_s=0.06$.

The mean and $rms$ velocity profiles of the `reference' ($C_s=0.05$ and $C_s=0.06$ for the highest $\Rey$) solutions obtained at these larger Reynolds numbers are reported in \reffig{HighReNTW}.
The change of the wall-normal and spanwise $rms$ velocity profiles with the Reynolds number is  minor.
The peak streamwise $rms$ velocity, associated to the large-scale streaks, increases with the Reynolds number and is attained nearer to the wall. 
All the $rms$ velocity components remain relatively large in the bulk of the flow.
These features are consistent with the observed behaviour of coherent large-scale motions  \cite[see e.g.][for the most recent DNS results]{Pirozzoli2014,Avsarkisov2014}.
This similarity is confirmed by the analysis of the flow fields associated to the large-scale coherent steady solutions reported in \reffig{UBHighRe}. 
This figure also illustrates how the eddy viscosity associated to the residual (small-scale) motions increases with the Reynolds number and is located mainly on the flanks of the low speed streak. 
The absolute levels of eddy viscosity are not huge even at $\Rey=2187$
where the maximum of $\nu_t$ does not exceed $90\%$ of the molecular
viscosity for those solutions but, contrary to the Navier-Stokes case,
the eddy viscosity is not spatially uniform and depends on the local
properties of the large-scale flow.

Having observed a clear relation between NBCW and coherent large-scale motions, it remains to verify if these solutions are also related to buffer-layer structures.
This question is addressed in appendix~B where it is shown that NBCW  do not converge to buffer layer structures if continued to higher $\Rey$ in minimal flow units with $L_x^+ \times L_z^+ = 250 \times 100$.

%%%%%%%%%%%%%%%%%%%%%%%%%%%%%%%%%%%%%%%%%%%%%%%%%%%%%%%%%%%%%%%%%%%%%%%%%%%%%%%
\section{Summary and discussion}
\label{sec:disc}

The main scope of this investigation has been to understand the origin of the coherent large-scale motions (LSM) which are numerically and experimentally observed in the turbulent plane Couette flow. 

In the first part of the study, the self-sustained nature of large-scale motions has been investigated using the over-damped large eddy simulation technique introduced by \cite{Hwang2010c,Hwang2011}.
The main results of the first part of the study can be summarized as follows:\\
(a) Reference large eddy simulations at $\Rey=2150$ (corresponding to  $\Retau=128$) in the \VLSMBOX\, ($L_x \times L_z =  132 h \times 12.5 h$) and with a sufficiently refined grid are able to capture the most important features of turbulent Couette flow, namely the near-wall cycle and the large and very large-scale motions (LSM \& VLSM) with characteristics sizes in good agreement with those found in the direct numerical simulations of \cite{Tsukahara2006}. \\
(b) Large-scale motions (LSM) do survive the quenching of the near-wall cycle at $\Rey=2150$ with properties very similar to those of the `natural' large-scale motions.  \\
(c) Large-scale motions also survive the quenching of the near-wall cycle in a `\LSMBOX'  with dimensions  $L_x \times L_z =  10.9 h \times 5.5 h$ typical of large-scale motions but where potentially active very-large scale motions of finite wavelength are excluded. 

These results further confirm the findings of \cite{Hwang2010b} for the turbulent pressure-driven channel that a self-sustained process is at work at large scale in wall-bounded shear flows.
The process is self-sustained because it does not rely on active motions at smaller or larger scales and its existence is in contrast with the current view that large-scale motions (LSM) ``are believed to be created by the vortex packets formed when multiple hairpin structures travel at the same convective velocity'' \citep[as summarized in the review paper by][]{Smits2011}.
Indeed, figures~\ref{fig:Streaks} and \ref{fig:sideview} clearly show that large-scale motions are still active even when the multiple smaller scale structures are removed from the scene.
It must be noted that the LSM of the over-damped simulations can be only qualitatively similar to the real ones because the near-wall self-sustained process has been artificially shut down in order to prove that LSM are self-sustained. It is therefore almost surprising that the LSM of the over-damped simulations still reproduce so many features of the `real' ones despite the very strong modification of the flow induced by the over-damping.

In the second part of the study the nature of the self-sustained large-scale motions in turbulent Couette flow has been further investigated by looking for steady state solutions of the filtered equations (i.e. of  the, possibly overdamped, LES) in a periodic domain with the typical  dimensions of the LSM (the \LSMBOX\, with $L_x \times L_z =  10.9 h \times 5.5 h$), which also coincide with those of the `optimum' domain considered by \cite{Waleffe2003} for the transitional Couette flow.
The main results can be summarized as follows:\\
(a) 
An upper and a lower branch of coherent large-scale steady solution of the filtered (LES) equations have been computed at $\Rey=750$ by edge-tracking and using Newton iterations.
These solutions have been continued in $C_s$ in the \LSMBOX. 
The upper and lower branches are connected via a saddle-node bifurcation at high $C_s$.
\\
(b) The continuation in $C_s$ also shows that both upper and lower branch solutions can be continued to the reference value $C_s=0.05$ and even to steady solutions of the Navier-Stokes equations obtained for $C_s=0$ up to $\Rey=1164$.\\
(c) Coherent steady upper-branch solutions at the reference value $C_s=0.05$ have been computed at Reynolds numbers up to $\Rey=2150$ using specific paths in the $\Rey-C_s$ plane.
The maximum amplitude of the associated large-scale streaks increases with the Reynolds number and its position approaches the wall on increasing $\Rey$. 
The eddy viscosity associated with these solutions also increases with $\Rey$. \\
(d) The solutions obtained for $C_s=0$ are shown to belong to the well known branch of  NBCW  solutions of the Navier-Stokes equations originating in a saddle-node bifurcation at $\Rey=127$ (and $C_s=0$) in the \LSMBOX.\\
(e) The continuation of the NBCW  upper branch solutions to high Reynolds numbers in a minimal flow unit ($L_x^+ \times L_z^+ = 250 \times 100$) does not converge to solutions consistent with typical buffer-layer structures, as shown in appendix~B.

This is, to the best of our knowledge, the first time that invariant `exact' solutions are computed for large-scale turbulent {\it filtered}, i.e. truly coherent, motions. 
These `exact' solutions of the LES equations, contrary to solutions of the Navier-Stokes equations take into account the effect of small scales only through their averaged effect.
The spatial and Reynolds number dependence of the eddy viscosity associated with the averaged (residual) small-scale motions is naturally embedded in the computed solutions. 
It is, in this way, possible to compute coherent large-scale steady solutions despite the fact that motions at smaller scale are unsteady and therefore concentrate on the interesting dynamics of the large-scale coherent solutions without the less relevant complications associated to motions at smaller scales. 

The features of the computed coherent steady state upper branch solutions are consistent with previous findings.
For instance, the fact that the maximum streak amplitude associated with these large-scale solutions increases with the Reynolds number is consistent with the fact that the large-scale peak in premultiplied spectra of turbulent wall-bounded flows also increases with Reynolds number  \citep[e.g.][]{Pirozzoli2014}.
It is also consistent with the fact that the maximum energy amplification of the most amplified streaks increases with the Reynolds number as predicted by the theoretical analyses of  \cite{Pujals2009}, \cite{Cossu2009}, \cite{Hwang2010,Hwang2010c} and \cite{Willis2010}.
That the position of the maximum streak amplitude approaches the wall for increasing $\Rey$ is also in agreement with previous findings \citep[see e.g.][]{Mathis2011}. 

Another interesting result is that the coherent large-scale solutions are connected to the NBCW  solutions of the Navier-Stokes equations and are relatively similar to them
when continued to higher Reynolds numbers in the \LSMBOX\, whose dimensions remain constant in outer units.
In previous studies the dynamics of NBCW  solutions has been discussed for low Reynolds numbers where the spanwise size of large-scale and buffer-layer streaks  are very similar. 
This ambiguity is removed here by considering sufficiently large Reynolds numbers where scale separation sets in between inner and outer layer structures and by showing that the 
NBCW  solutions do not converge to typical buffer-layer structures when continued to higher $\Rey$ in a minimal flow unit.
NBCW  solutions can therefore be considered as the `precursors' of large-scale motions.

This study is a second step, after that of \cite{Hwang2010b,Hwang2011}, towards a `dynamical systems'  understanding of the large-scale dynamics in fully developed turbulent shear flows. 
The key ingredient is to model small-scale motions and to only resolve large-scale motions in order to compute invariant solutions.
Much work remains to be done to assess the relevance of coherent invariant solutions of the filtered equations to the prediction of turbulent flow dynamics. 
Interesting questions left for future work are, for instance, to understand if other steady or periodic solutions of the LES equations can be computed and to know how much time turbulent solutions of  large-eddy simulations spend in the neighbourhood of exact coherent large-scale solutions \citep[in the same spirit of the Navier-Stokes computations of][]{Kerswell2007,Schneider2007}.
One could indeed hope that the increase in effective viscosity in fully developed turbulent large-scale flows, could lead to a `simpler' dynamics of large-scale motions where a few invariant solutions are sufficient to capture essential features of the flow as in the case of transitional flows \citep[see e.g.][]{Kawahara2001}.

As the static Smagorinsky model used in this work is very crude, there certainly is room for improvement in the modelling of small-scale dynamics to compute self-sustained large-scale coherent motions more accurately.
However, we believe that meaningful results can be found in other canonical wall-bounded shear flows using this over-damped LES technique even in its relatively primitive current form. 
The methods used in this study could also help to shed light on different problems involving shear flows, as for instance the very large Reynolds and/or magnetic Reynolds number behaviour of self-sustained processes in Keplerian shear flow \citep{Rincon2007b,Rincon2007,Rincon2008,Herault2011,Riols2013}, with possible practical applications to the modelling of the dynamics of astrophysical accretion disks \citep{Riols2015}.

%%%%%%%%%%%%%%%%%%%%%%%%%%%%%%%%%%%%%%%%%%%%%%%%%%%%%%%%%%%%%%%%%%%%%%%%%%%%%%%
%%%%%%%%%%%%%%%%%%%%%%%%%%%%%%%%%%%%%%%%%%%%%%%%%%%%%%%%%%%%%%%%%%%%%%%%%%%%%%%
\appendix

% %=====================================================================================================
\section{Reference and overdamped LES in the LSM-BOX}
\label{sec:LESLSMBOX}

\begin{figure}
%\vspace{-8mm}
\centering
\includegraphics[width=0.80\textwidth]{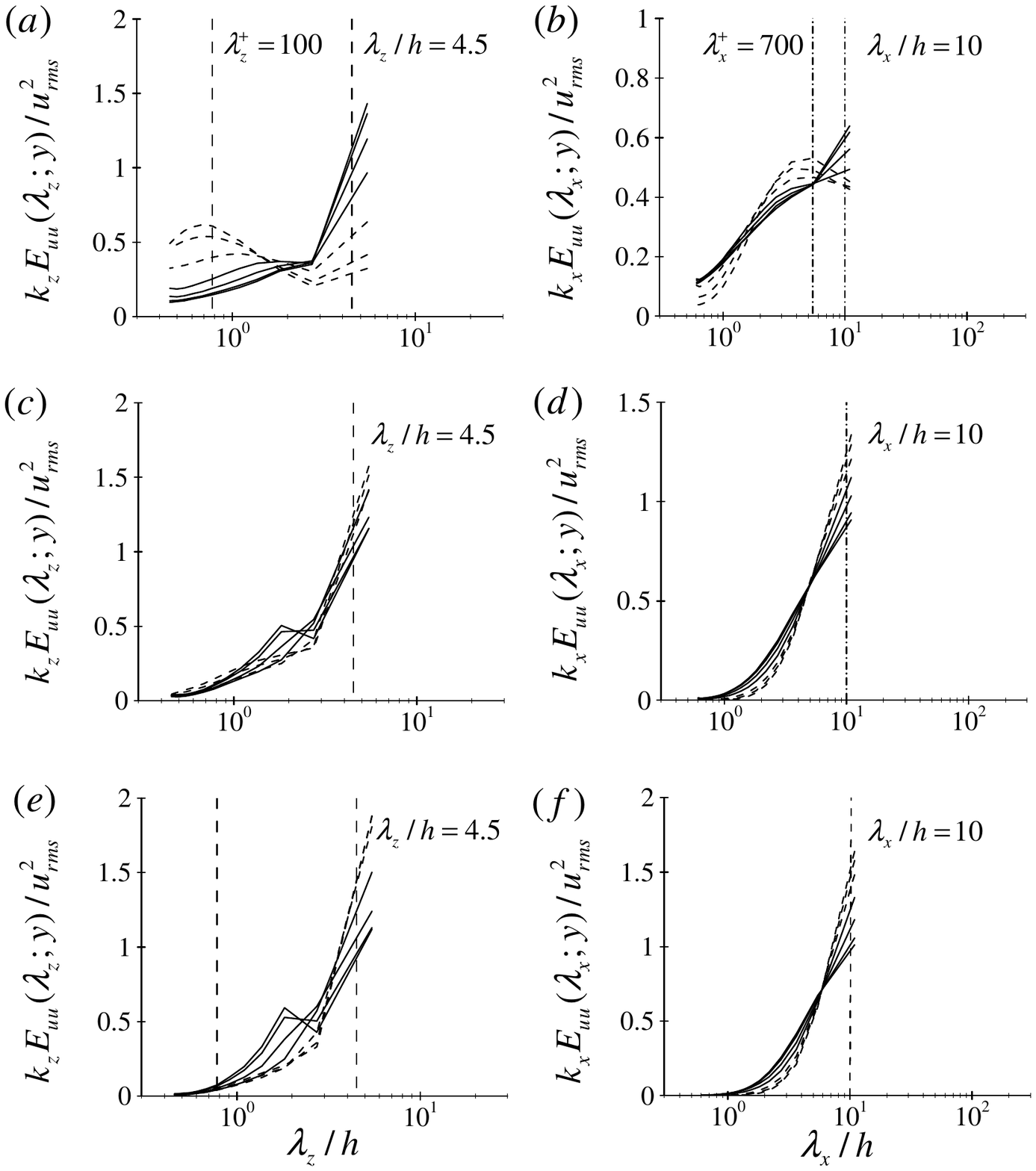} 
%\vspace{-45mm}
\caption{Premultiplied one-dimensional spanwise  $(a,c,e)$ and streamwise $(b,d,f)$ spectra of the streamwise velocity in the \LSMBOX\, obtained for increasing values of the Smagorinsky constant: $(a,b)$ $C_s=0.05$ (reference value); $(c,d)$ $C_s=0.14$; $(e,f)$ $C_s=0.18$.}
\label{fig:figa1}
\end{figure}

A series of simulations have been performed in the \LSMBOX\, with dimensions  
$L_x \times L_z =  10.9 h \times 5.5 h$  typical of large-scale motions in order to verify if these motions can self-sustain in the absence of the potentially active very-large scale motions that were captured in the \VLSMBOX~discussed in \refsec{SSPLSM}.
The number of grid points used in this box (see \reftab{BOXES}) is chosen so as to keep the same grid spacing as in the \VLSMBOX.
Results are obtained at the same Reynolds number ($\Rey=2150$, corresponding to $\Retau=128$) and for the same values of $C_s$ considered in \refsec{SSPLSM} for the \VLSMBOX.
The spanwise and streamwise premultiplied spectra of the streamwise velocity component are shown in \reffig{figa1}, which is the analogue of \reffig{PremSpec} of \refsec{SSPLSM}. 

For the Reynolds number considered, the \LSMBOX\, (whose dimensions in inner units are $L_x^+ \times L_z^+=1395 \times 678$) is large enough to accommodate many near-wall structures. 
Indeed, at the reference value $C_s=0.05$, the near-wall peaks at $\lambda_z^+ \simeq 100$ and at $\lambda_x^+ \simeq 700$ are clearly visible in the spanwise and the streamwise spectra,  respectively reported in figures \ref{fig:figa1}$a$ and  \ref{fig:figa1}$b$. 
The spectral energy associated with the outer structures appears to be highly concentrated at the largest wavelengths allowed in the given computational domain which correspond to large-scale motions (LSM). 
As already mentioned, very-large scale motions (of finite streamwise wavelength) are out of the picture in the \LSMBOX. 
On increasing $C_s$, the near-wall peaks are gradually quenched as in the spectra obtained in the \VLSMBOX\, (see \reffig{PremSpec}). 
The outer structures in the confined domain appear to be well isolated at $C_s=0.14$ (figures \ref{fig:figa1}$c$ and $d$) and persist even for the further increase of $C_s$ (figures \ref{fig:figa1}$e$ and $f$).
These results definitely confirm that the large-scale motions (LSM) are self-sustained in the absence not only of near-wall active processes but also in the absence of potentially active very-large scale motions (VLSM) of finite wavelength.

%=====================================================================================================
\section{Continuation of the NBCW solutions in a MFU}
\label{sec:BULABOX}

\begin{figure}
 \centering
% ~ \hspace{-8mm}
 \includegraphics[width=0.32\columnwidth,height=0.35\columnwidth]{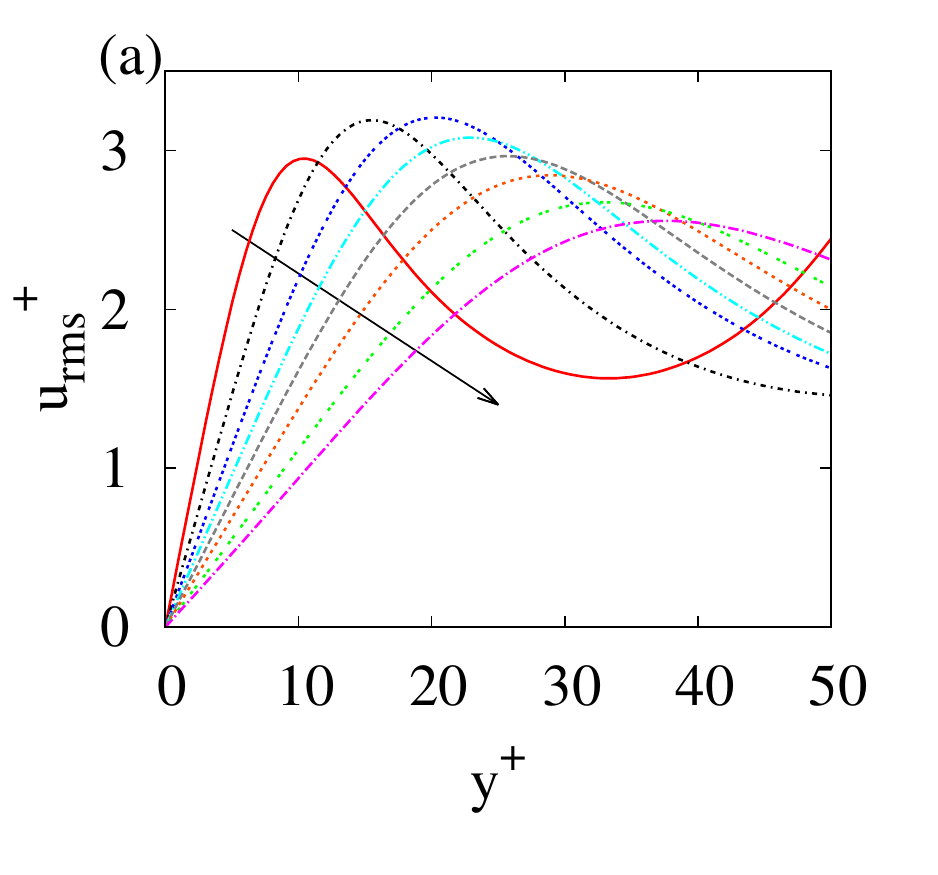} %\hspace{-8mm}
 \includegraphics[width=0.32\columnwidth,height=0.35\columnwidth]{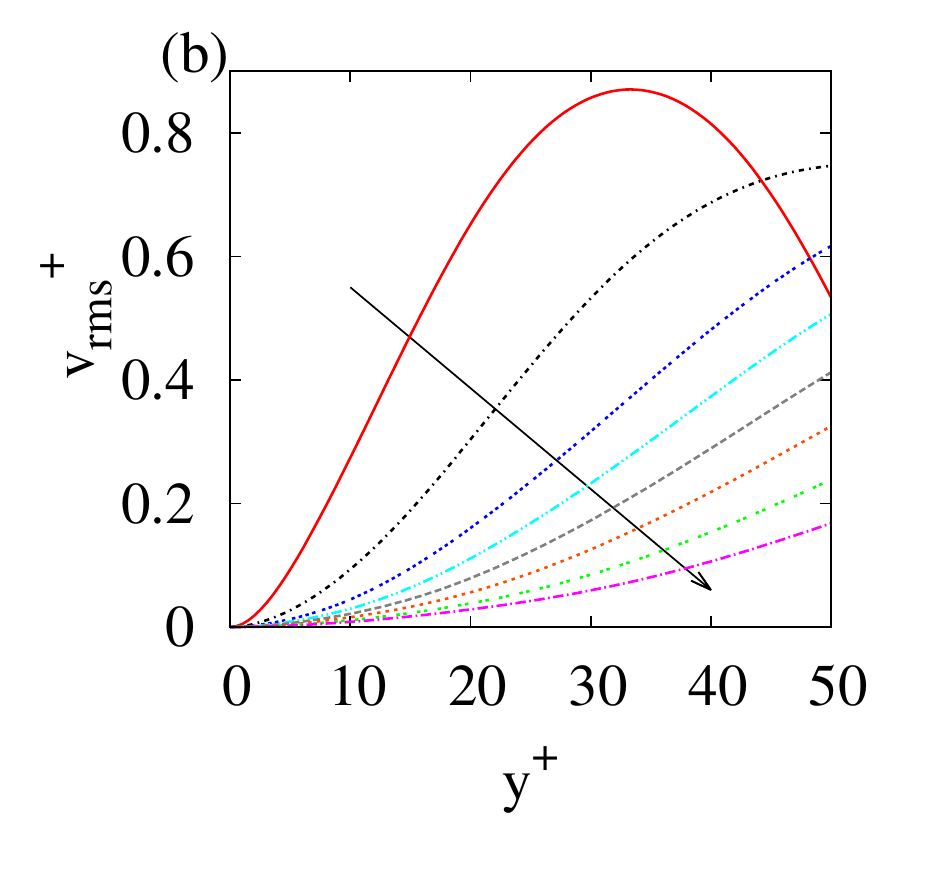} %\hspace{-8mm}
 \includegraphics[width=0.32\columnwidth,height=0.35\columnwidth]{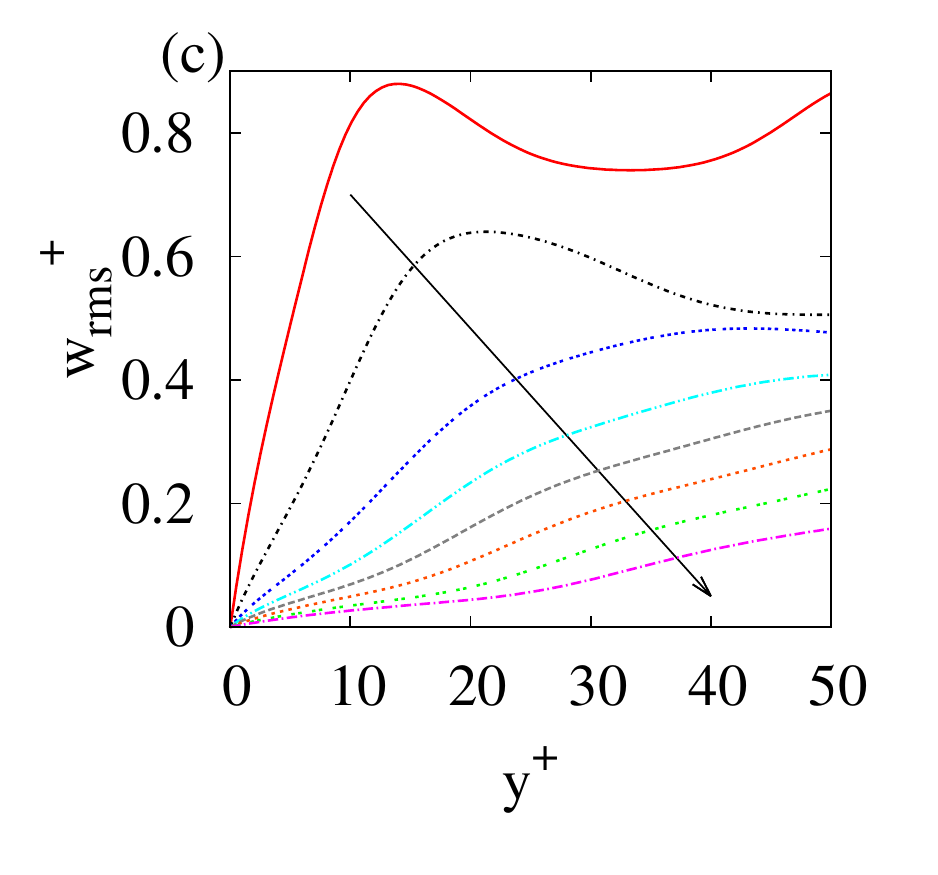}
 \caption{$rms$ velocity profiles of the
nonlinear steady solutions obtained by continuation to higher
Reynolds numbers of the NBCW  solutions in minimal flow units with
$L_x^+=250$ and $L_z^+=100$. The data corresponds to $\Rey=400, 750, 1100, 1200,\ldots, 1500, 1600$,
with arrows indicating the increase in Reynolds number. 
}
 \label{fig:NearWallSols}
\end{figure}

In this appendix we investigate if NBCW  solutions can be continued into near-wall structures when the Reynolds number is sufficiently increased in minimal flow units (MFU). 
This issue is addressed in the case of the Navier-Stokes equations ($C_s=0$).  

The main characteristic of the near-wall streaky structures  is that their typical spanwise and streamwise lengths decrease when the Reynolds number is increased, but remain constant when expressed in wall units.
Continuation of these solutions to higher Reynolds numbers should therefore be made keeping the box dimension fixed {\it in wall units}. 
As $L_z^+ = \Retau L_z / h$, when $\Retau$ is increased, $L_z$ must be decreased in order to keep constant $L_z^+$, and similarly for $L_x$. 
The approximate relation $\Retau \approx 0.054\,\Rey+11.22$ which fits
well the data of a series of direct numerical simulations performed in
the \LSMBOX~for $\Rey \in [400,2150]$, is used to determine the box
size. Attention must be paid to the grid resolution in the wall-normal direction because the size of the buffer-layer region, where near-wall streaks reside, also decreases when the Reynolds number is increased. 
In order to achieve a sufficient accuracy in the wall-normal
direction, keeping the state vector size manageable for the Newton
iterations, the time integrations required to perform the computation
of the nonlinear solutions presented here have been performed using
the {\tt channelflow} code \cite{Gibson2008} which uses a Chebyshev
spectral discretization in $y$, instead of the second-order finite
difference discretization of {\tt diablo}.  
This code can be used here because we do not consider large eddy simulations in this appendix.
For these computations we use $N_x \times N_y \times N_z = 32 \times 65 \times 32$ points.
We have verified that the results do not change in any noticeable way when the number of points in the wall-normal direction is increased to $N_y=95$ for the highest considered Reynolds numbers.

The NBCW  upper branch solution at $\Rey=400$
is used as a starting point of the continuation.
At $\Rey=400$ the size of the \LSMBOX, in inner units, is $L_z^+ \approx 180$ and $L_x^+ \approx 360$.
As a preliminary step, this solution is continued in the parameters $L_x$ and $L_z$ to achieve $L_z^+ \approx 100$ and $L_x^+ \approx 200$ (which in external units corresponds to $L_z=2.6 h$ and $L_x=5.51 h$ \citep[see also][]{Gibson2008}). 
This steady solution is then continued by increasing the Reynolds number in small steps, changing the box size so as to keep it constant in inner units to $L_z^+ = 100$ and $L_x^+ = 250$ and predicting the initial guess on the velocity fields expressed in inner units.
The Newton-based iterations are performed using {\tt peanuts}.

The velocity profiles of the converged solutions are reported in \reffig{NearWallSols} for selected Reynolds numbers and they appear to always belong to the same branch. 
The results show that the solutions continued in the near-wall minimal flow unit do not converge to near-wall structures.
For instance, the $y^+$ position of the maximum of the $rms$  velocity
profiles increases when $\Retau$ is increased, while it should
instead remain constant, as found by \cite{Jimenez1991}, in minimum
flow unit simulations \citep[see also][]{Hwang2013b}. 
The NBCW  solutions do not therefore seem to be connected to near-wall structures, at least in plane Couette flow, even if continued in minimal flow units, i.e. periodic domains which remain constant in inner units while increasing the Reynolds number.

%%%%%%%%%%%%%%%%%%%%%%%%%%%%%%%%%%%%%%%%%%%%%%%%%%%%%%%%%%%%%%%%%%%%%%%%%%%%%%%%%%%%%%%%%%%%%%%%%%%%%%%%%%%%%%%%%%%%%%%%%%

%%%%%%%%%%%%%%%%%%%%%%%%%%%%%%%%%%%%%%%%%%%%%%%%%%%%%%%%%%%%%%%%%%%%%%%%%%%%%%%%%%%%%%%%%%%%%%%%%%%%%%%%%%%%%%%%%%%
\begin{acknowledgments}
The use of the {\tt channelflow} and {\tt diablo} codes and financial support from PRES Universit\'e de Toulouse and R\'egion Midi-Pyr\'en\'ees are kindly acknowledged.
\end{acknowledgments}

% \bibliographystyle{jfm}
 %\bibliography{carlo3,rawat,missingbib}

\begin{thebibliography}{92}
\expandafter\ifx\csname natexlab\endcsname\relax\def\natexlab#1{#1}\fi

\bibitem[Adrian(2007)]{Adrian2007}
{\sc Adrian, R.~J.} 2007 Hairpin vortex organization in wall turbulence. {\em
  Phys. Fluids.\/} {\bf 19}, 041301.

\bibitem[Adrian {\em et~al.\/}(2000)Adrian, Meinhart \& Tomkins]{Adrian2000}
{\sc Adrian, R.~J., Meinhart, C.~D. \& Tomkins, C.~D.} 2000 Vortex organization
  in the outer region of the turbulent boundary layer. {\em J. Fluid. Mech.\/}
  {\bf 422}, 1--54.

\bibitem[del {\'A}lamo \& Jim{\'e}nez(2003)]{delAlamo2003}
{\sc del {\'A}lamo, J.~C. \& Jim{\'e}nez, J.} 2003 {Spectra of the very large
  anisotropic scales in turbulent channels}. {\em Phys. Fluids\/} {\bf 15},
  L41.

\bibitem[del \'{A}lamo \& Jim\'{e}nez(2006)]{delAlamo2006}
{\sc del \'{A}lamo, J.~C. \& Jim\'{e}nez, J.} 2006 Linear energy amplification
  in turbulent channels. {\em J. Fluid Mech.\/} {\bf 559}, 205--213.

\bibitem[Artuso {\em et~al.\/}(1990)Artuso, Aurell \& Cvitanovic]{Artuso1990}
{\sc Artuso, R, Aurell, E \& Cvitanovic, P} 1990 Recycling of strange sets: I.
  cycle expansions. {\em Nonlinearity\/} {\bf 3}, 325.

\bibitem[Avsarkisov {\em et~al.\/}(2014)Avsarkisov, Hoyas, Oberlack \&
  Garcia-Galache]{Avsarkisov2014}
{\sc Avsarkisov, V., Hoyas, S., Oberlack, M. \& Garcia-Galache, J.P.} 2014
  Turbulent plane {C}ouette flow at moderately high {R}eynolds number. {\em J.
  Fluid Mech.\/} {\bf 751}, R1--1--R2--9.

\bibitem[Balay {\em et~al.\/}(2011)Balay, Brown, Buschelman, Gropp, Kaushik,
  Knepley, McInnes, Smith \& Zhang]{petsc}
{\sc Balay, Satish, Brown, Jed, Buschelman, Kris, Gropp, William~D., Kaushik,
  Dinesh, Knepley, Matthew~G., McInnes, Lois~Curfman, Smith, Barry~F. \& Zhang,
  Hong} 2011 {PETSc} {W}eb page. {\texttt{http://www.mcs.anl.gov/petsc}}.

\bibitem[Bewley {\em et~al.\/}(2001)Bewley, Moin \& Temam]{Bewley2001}
{\sc Bewley, T.~R., Moin, P. \& Temam, R.} 2001 {DNS}-based predictive control
  of turbulence: an optimal benchmark for feedback algorithms. {\em J. Fluid
  Mech.\/} {\bf 447}, 179--225.

\bibitem[Blackwelder \& Kovasznay(1972)]{Blackwelder1972}
{\sc Blackwelder, R.~F. \& Kovasznay, L. S.~G.} 1972 {Time scales and
  correlations in a turbulent boundary layer}. {\em Phys. Fluids.\/} {\bf 15},
  1545--1554.

\bibitem[Chandler \& Kerswell(2013)]{Chandler2013}
{\sc Chandler, G~J \& Kerswell, R~R} 2013 Invariant recurrent solutions
  embedded in a turbulent two-dimensional {K}olmogorov flow. {\em J. Fluid
  Mech.\/} {\bf 722}, 554--595.

\bibitem[Clever \& Busse(1997)]{Clever1997}
{\sc Clever, RM \& Busse, FH} 1997 Tertiary and quaternary solutions for plane
  {C}ouette flow. {\em J. Fluid Mech.\/} {\bf 344}, 137--153.

\bibitem[Clever \& Busse(1992)]{Clever1992}
{\sc Clever, R.~M. \& Busse, F.~H.} 1992 Three-dimensional convection in a
  horizontal fluid layer subjected to a constant shear. {\em J. Fluid Mech.\/}
  {\bf 234}, 511--527.

\bibitem[Corrsin \& Kistler(1954)]{Corrsin1954}
{\sc Corrsin, S. \& Kistler, A.~L.} 1954 The free-stream boundaries of
  turbulent flows. {\em Technical Note\/} {\bf 3133}, 120--130, nACA.

\bibitem[Cossu {\em et~al.\/}(2011)Cossu, L., Bagheri \& Henningson]{Cossu2011}
{\sc Cossu, C., L., Brandt, Bagheri, S. \& Henningson, D.~S..} 2011 Secondary
  threshold amplitudes for sinuous streak breakdown. {\em Phys. Fluids\/} {\bf
  23}, 074103.

\bibitem[Cossu {\em et~al.\/}(2009)Cossu, Pujals \& Depardon]{Cossu2009}
{\sc Cossu, C., Pujals, G. \& Depardon, S.} 2009 Optimal transient growth and
  very large scale structures in turbulent boundary layers. {\em J. Fluid
  Mech.\/} {\bf 619}, 79--94.

\bibitem[Deardorff(1970)]{Deardorff1970}
{\sc Deardorff, P.~E.} 1970 A numerical study of three-dimensional turbulent
  channel flow at large {R}eynolds numbers. {\em J. Fluid Mech.\/} {\bf 41},
  453--480.

\bibitem[Dennis \& Nickels(2011)]{Dennis2011b}
{\sc Dennis, D. J.~C. \& Nickels, T.~B.} 2011 Experimental measurement of
  large-scale three-dimensional structures in a turbulent boundary layer.
  {P}art 2. long structures. {\em J. Fluid Mech.\/} {\bf 673}, 218--244.

\bibitem[Eitel-Amor {\em et~al.\/}(2015)Eitel-Amor, {\"O}rl{\"u}, Schlatter \&
  Flores]{Eitel2015}
{\sc Eitel-Amor, G, {\"O}rl{\"u}, R, Schlatter, P \& Flores, O} 2015 Hairpin
  vortices in turbulent boundary layers. {\em Phys. Fluids\/} {\bf 27}~(2),
  025108.

\bibitem[Ellingsen \& Palm(1975)]{Ellingsen1975}
{\sc Ellingsen, T. \& Palm, E.} 1975 Stability of linear flow. {\em Phys.
  Fluids\/} {\bf 18}, 487.

\bibitem[Falco(1977)]{Falco1977}
{\sc Falco, R.~E.} 1977 Coherent motions in the outer region of turbulent
  boundary layers. {\em Phys. Fluids\/} {\bf 20}, S124--S132.

\bibitem[Flores \& Jim\'enez(2006)]{Flores2006}
{\sc Flores, O. \& Jim\'enez, J.} 2006 {Effect of wall-boundary disturbances on
  turbulent channel flows}. {\em J. Fluid Mech.\/} {\bf 566}, 357--376.

\bibitem[Flores \& Jim{\'e}nez(2010)]{Flores2010}
{\sc Flores, O. \& Jim{\'e}nez, J.} 2010 {Hierarchy of minimal flow units in
  the logarithmic layer}. {\em Phys. Fluids\/} {\bf 22}, 071704.

\bibitem[Flores {\em et~al.\/}(2007)Flores, Jim\'enez \& del
  {\'A}lamo]{Flores2007}
{\sc Flores, O., Jim\'enez, J. \& del {\'A}lamo, J.C.} 2007 {Vorticity
  organization in the outer layer of turbulent channels with disturbed walls}.
  {\em J. Fluid Mech.\/} {\bf 591}, 145--154.

\bibitem[Gibson {\em et~al.\/}(2008)Gibson, Halcrow \& Cvitanovic]{Gibson2008}
{\sc Gibson, J.~F., Halcrow, J. \& Cvitanovic, P.} 2008 Visualizing the
  geometry of state space in plane {C}ouette flow. {\em J. Fluid Mech.\/} {\bf
  611}, 107--130.

\bibitem[Guala {\em et~al.\/}(2006)Guala, Hommema \& Adrian]{Guala2006}
{\sc Guala, M., Hommema, S.~E. \& Adrian, R.~J.} 2006 Large-scale and
  very-large-scale motions in turbulent pipe flow. {\em J. Fluid Mech.\/} {\bf
  554}, 521--541.

\bibitem[Hamilton {\em et~al.\/}(1995)Hamilton, Kim \& Waleffe]{Hamilton1995}
{\sc Hamilton, J.M., Kim, J. \& Waleffe, F.} 1995 {Regeneration mechanisms of
  near-wall turbulence structures}. {\em J. Fluid Mech\/} {\bf 287}, 317--348.

\bibitem[H\"{a}rtel \& Kleiser(1998)]{Hartel1998}
{\sc H\"{a}rtel, C. \& Kleiser, L.} 1998 Analysis and modelling of
  subgrid-scale motions in near-wall turbulence. {\em J. Fluid Mech\/} {\bf
  356}, 327--352.

\bibitem[Head \& Bandyopadhay(1981)]{Head1981}
{\sc Head, M.~R. \& Bandyopadhay, P.} 1981 New aspects of turbulent
  boundary-layer structure. {\em J. Fluid Mech\/} {\bf 107}, 297--338.

\bibitem[Herault {\em et~al.\/}(2011)Herault, Rincon, Cossu, Lesur, Ogilvie \&
  Longaretti]{Herault2011}
{\sc Herault, J., Rincon, F., Cossu, C., Lesur, G., Ogilvie, G.~I. \&
  Longaretti, P.Y.} 2011 Periodic magnetorotational dynamo action as a
  prototype of nonlinear magnetic field generation in shear flows. {\em Phys.
  Rev. E\/} {\bf 84}, 036321.

\bibitem[Hutchins \& Marusic(2007{\natexlab{{\em a\/}}})]{Hutchins2007}
{\sc Hutchins, N. \& Marusic, I.} 2007{\natexlab{{\em a\/}}} {Evidence of very
  long meandering features in the logarithmic region of turbulent boundary
  layers}. {\em J. Fluid Mech.\/} {\bf 579}, 1--28.

\bibitem[Hutchins \& Marusic(2007{\natexlab{{\em b\/}}})]{Hutchins2007b}
{\sc Hutchins, N. \& Marusic, I.} 2007{\natexlab{{\em b\/}}} Large-scale
  influences in near-wall turbulence. {\em Phil. Trans. R. Soc. A\/} {\bf 365},
  647--664.

\bibitem[Hwang(2013)]{Hwang2013b}
{\sc Hwang, Y} 2013 Near-wall turbulent fluctuations in the absence of wide
  outer motions. {\em J. Fluid Mech.\/} {\bf 723}, 264--288.

\bibitem[Hwang(2015)]{Hwang2015}
{\sc Hwang, Y} 2015 Statistical structure of self-sustaining attached eddies in
  turbulent channel flow. {\em J. Fluid Mech.\/} {\bf 767}, 254--289.

\bibitem[Hwang \& Cossu(2010{\natexlab{{\em a\/}}})]{Hwang2010}
{\sc Hwang, Y. \& Cossu, C.} 2010{\natexlab{{\em a\/}}} Amplification of
  coherent streaks in the turbulent {C}ouette flow: an input-output analysis at
  low {R}eynolds number. {\em J. Fluid Mech.\/} {\bf 643}, 333--348.

\bibitem[Hwang \& Cossu(2010{\natexlab{{\em b\/}}})]{Hwang2010c}
{\sc Hwang, Y. \& Cossu, C.} 2010{\natexlab{{\em b\/}}} Linear non-normal
  energy amplification of harmonic and stochastic forcing in turbulent channel
  flow. {\em J. Fluid Mech.\/} {\bf 664}, 51--73.

\bibitem[Hwang \& Cossu(2010{\natexlab{{\em c\/}}})]{Hwang2010b}
{\sc Hwang, Y. \& Cossu, C.} 2010{\natexlab{{\em c\/}}} {Self-sustained process
  at large scales in turbulent channel flow.} {\em Phys. Rev. Lett.\/} {\bf
  105}~(4), 044505.

\bibitem[Hwang \& Cossu(2011)]{Hwang2011}
{\sc Hwang, Y. \& Cossu, C.} 2011 Self-sustained processes in the logarithmic
  layer of turbulent channel flows. {\em Phys. Fluids\/} {\bf 23}, 061702.

\bibitem[Itano \& Toh(2001)]{Itano2001}
{\sc Itano, T. \& Toh, S.} 2001 The dynamics of bursting process in wall
  turbulence. {\em J. Phys. Soc. Jpn.\/} {\bf 70}, 703--716.

\bibitem[Jeong {\em et~al.\/}(1997)Jeong, Hussain, Schoppa \& Kim]{Jeong1997}
{\sc Jeong, J., Hussain, F., Schoppa, W. \& Kim, J.} 1997 Coherent structures
  near the wall in a turbulent channel flow. {\em J. Fluid Mech.\/} {\bf 332},
  185--214.

\bibitem[Jim{\'e}nez \& Moin(1991)]{Jimenez1991}
{\sc Jim{\'e}nez, J. \& Moin, P.} 1991 {The minimal flow unit in near-wall
  turbulence}. {\em J. Fluid Mech.\/} {\bf 225}, 213--240.

\bibitem[Kawahara \& Kida(2001)]{Kawahara2001}
{\sc Kawahara, G. \& Kida, S.} 2001 {Periodic motion embedded in plane
  {C}ouette turbulence: regeneration cycle and burst}. {\em J. Fluid Mech.\/}
  {\bf 449}, 291--300.

\bibitem[Kerswell \& Tutty(2007)]{Kerswell2007}
{\sc Kerswell, R.~R. \& Tutty, O.R.} 2007 Recurrence of travelling waves in
  transitional pipe flow. {\em J.\ Fluid Mech.\/} {\bf 584}, 69--102.

\bibitem[Kim {\em et~al.\/}(1987)Kim, Moin \& Moser]{Kim1987}
{\sc Kim, J, Moin, P \& Moser, R} 1987 Turbulence statistics in fully developed
  channel flow at low {R}eynolds number. {\em J. Fluid Mech.\/} {\bf 177},
  133--166.

\bibitem[Kim \& Adrian(1999)]{Kim1999}
{\sc Kim, K.~C. \& Adrian, R.} 1999 Very large-scale motion in the outer layer.
  {\em Phys. Fluids\/} {\bf 11}~(2), 417--422.

\bibitem[Kim \& Menon(1999)]{Menon1999}
{\sc Kim, W.W \& Menon, S.} 1999 An unsteady incompressible navier-stokes
  solver for large eddy simulation of turbulent flows. {\em Int. J. Numer.
  Meth. Fluids\/} {\bf 31}, 983--1017.

\bibitem[Kitoh {\em et~al.\/}(2005)Kitoh, Nakabayashi \& Nishimura]{Kitoh2005}
{\sc Kitoh, O., Nakabayashi, K. \& Nishimura, F.} 2005 Experimental study on
  mean velocity and turbulence characteristics of plane {C}ouette flow:
  low-{R}eynolds-number effects and large longitudinal vortical structures.
  {\em J. Fluid Mech.\/} {\bf 539}, 199.

\bibitem[Kitoh \& Umeki(2008)]{Kitoh2008}
{\sc Kitoh, O. \& Umeki, M.} 2008 Experimental study on large-scale streak
  structure in the core region of turbulent plane {C}ouette flow. {\em Phys.
  Fluids\/} {\bf 20}, 025107.

\bibitem[Kline {\em et~al.\/}(1967)Kline, {R}eynolds, Schraub \&
  Runstadler]{Kline1967}
{\sc Kline, S.~J., {R}eynolds, W.~C., Schraub, F.~A. \& Runstadler, P.~W.} 1967
  The structure of turbulent boundary layers. {\em J. Fluid Mech.\/} {\bf 30},
  741--773.

\bibitem[Komminaho {\em et~al.\/}(1996)Komminaho, Lundbladh \&
  Johansson]{Komminaho1996}
{\sc Komminaho, J., Lundbladh, A. \& Johansson, A.~V.} 1996 Very large
  structures in plane turbulent {C}ouette flow. {\em J. Fluid Mech.\/} {\bf
  320}, 259--285.

\bibitem[Kovasznay {\em et~al.\/}(1970)Kovasznay, Kibens \&
  Blackwelder]{Kovasznay1970}
{\sc Kovasznay, L. S.~G., Kibens, V. \& Blackwelder, R.~F.} 1970 {Large-scale
  motion in the intermittent region of a turbulent boundary layer}. {\em J.
  Fluid Mech.\/} {\bf 41}, 283--325.

\bibitem[Landahl(1980)]{Landahl1980}
{\sc Landahl, M.~T.} 1980 A note on an algebraic instability of inviscid
  parallel shear flows. {\em J. Fluid Mech.\/} {\bf 98}, 243--251.

\bibitem[Landahl(1990)]{Landahl1990}
{\sc Landahl, M.~T.} 1990 On sublayer streaks. {\em J. Fluid Mech.\/} {\bf
  212}, 593--614.

\bibitem[Lee \& Kim(1991)]{Lee1991}
{\sc Lee, M.~J. \& Kim, J.} 1991 The structure of turbulence in a simulated
  plane {C}ouette flow. In {\em Eighth Symp. on Turbulent Shear Flow\/}, pp.
  5.3.1--5.3.6. Tech. University of Munich, Sept. 9-11.

\bibitem[Mason \& Callen(1986)]{Mason1986}
{\sc Mason, PJ \& Callen, NS} 1986 {On the magnitude of the subgrid-scale eddy
  coefficient in large-eddy simulations of turbulent channel flow}. {\em J.
  Fluid Mech.\/} {\bf 162}, 439--462.

\bibitem[Mathis {\em et~al.\/}(2009)Mathis, Hutchins \& Marusic]{Mathis2009}
{\sc Mathis, R., Hutchins, N. \& Marusic, I.} 2009 Large-scale amplitude
  modulation of the small-scale structures in turbulent boudnary layers. {\em
  J. Fluid. Mech.\/} {\bf 628}, 311--337.

\bibitem[Mathis {\em et~al.\/}(2011)Mathis, Hutchins \& Marusic]{Mathis2011}
{\sc Mathis, R., Hutchins, N. \& Marusic, I.} 2011 A predictive inner--outer
  model for streamwise turbulence statistics in wall-bounded flows. {\em J.
  Fluid. Mech.\/} {\bf 681}, 537--566.

\bibitem[Moffatt(1967)]{Moffatt1967}
{\sc Moffatt, H.~K.} 1967 {The interaction of turbulence with strong wind
  shear}. In {\em Proc. URSI-IUGG Coloq. on Atoms. Turbulence and Radio Wave
  Propag.\/} (ed. A.M. Yaglom \& V.~I. Tatarsky), pp. 139--154. Moscow: Nauka.

\bibitem[Nagata(1990)]{Nagata1990}
{\sc Nagata, M.} 1990 Three-dimensional finite-amplitude solutions in plane
  {C}ouette flow: bifurcation from infinity. {\em J. Fluid Mech.\/} {\bf 217},
  519--527.

\bibitem[Park {\em et~al.\/}(2011)Park, Hwang \& Cossu]{Park2011}
{\sc Park, J., Hwang, Y. \& Cossu, C.} 2011 {On the stability of large-scale
  streaks in turbulent {C}ouette and Poiseulle flows}. {\em C. R. M{\'e}c.\/}
  {\bf 339}, 1--5.

\bibitem[Perry \& Chong(1982)]{Perry1982}
{\sc Perry, A.~E. \& Chong, M.~S.} 1982 On the mechanism of turbulence. {\em J.
  Fluid Mech.\/} {\bf 119}, 173--217.

\bibitem[Pirozzoli {\em et~al.\/}(2014)Pirozzoli, Bernardini \&
  Orlandi]{Pirozzoli2014}
{\sc Pirozzoli, S/, Bernardini, M. \& Orlandi, P.} 2014 Turbulence statistics
  in {C}ouette flow at high {R}eynolds number. {\em J. Fluid Mech.\/} {\bf
  758}, 327--343.

\bibitem[Pope(2000)]{Pope2000}
{\sc Pope, S.~B.} 2000 {\em Turbulent flows\/}. Cambridge, UK: Cambridge U.
  Press.

\bibitem[Pujals {\em et~al.\/}(2010)Pujals, Cossu \& Depardon]{Pujals2010b}
{\sc Pujals, G., Cossu, C. \& Depardon, S.} 2010 Forcing large-scale coherent
  streaks in a zero pressure gradient turbulent boundary layer. {\em J.
  Turb.\/} {\bf 11}~(25), 1--13.

\bibitem[Pujals {\em et~al.\/}(2009)Pujals, Garc\'{\i}a-Villalba, Cossu \&
  Depardon]{Pujals2009}
{\sc Pujals, G., Garc\'{\i}a-Villalba, M., Cossu, C. \& Depardon, S.} 2009 A
  note on optimal transient growth in turbulent channel flows. {\em Phys.
  Fluids\/} {\bf 21}, 015109.

\bibitem[Rawat(2014)]{Rawat2014b}
{\sc Rawat, S.} 2014 Coherent dynamics of large-scale turbulent motions. PhD
  thesis, Universit\'e de Toulouse.

\bibitem[Reddy {\em et~al.\/}(1998)Reddy, Schmid, Baggett \&
  Henningson]{Reddy1998}
{\sc Reddy, S.~C., Schmid, P.~J., Baggett, J.~S. \& Henningson, D.~S.} 1998 On
  the stability of streamwise streaks and transition thresholds in plane
  channel flows. {\em J. Fluid Mech.\/} {\bf 365}, 269--303.

\bibitem[{R}eynolds \& Hussain(1972)]{Reynolds1972}
{\sc {R}eynolds, W.~C. \& Hussain, A. K. M.~F.} 1972 {The mechanics of an
  organized wave in turbulent shear flow. Part 3. Theoretical models and
  comparisons with experiments}. {\em J. Fluid Mech.\/} {\bf 54}~(02),
  263--288.

\bibitem[Rincon {\em et~al.\/}(2007{\natexlab{{\em a\/}}})Rincon, Ogilvie \&
  Proctor]{Rincon2007b}
{\sc Rincon, F., Ogilvie, GI \& Proctor, MRE} 2007{\natexlab{{\em a\/}}}
  {Self-Sustaining Nonlinear Dynamo Process in Keplerian Shear Flows}. {\em
  Phys. Rev. Lett.\/} {\bf 98}~(25), 254502.

\bibitem[Rincon {\em et~al.\/}(2007{\natexlab{{\em b\/}}})Rincon, Ogilvie \&
  Cossu]{Rincon2007}
{\sc Rincon, F., Ogilvie, G.~I. \& Cossu, C.} 2007{\natexlab{{\em b\/}}} On
  self-sustaining processes in {R}ayleigh-stable rotating plane {C}ouette flows
  and subcritical transition to turbulence in accretion disks. {\em Astron. \&
  Astrophys\/} {\bf 463}, 817--832.

\bibitem[Rincon {\em et~al.\/}(2008)Rincon, Ogilvie, Proctor \&
  Cossu]{Rincon2008}
{\sc Rincon, F., Ogilvie, G.~I., Proctor, M. R.~E. \& Cossu, C.} 2008
  Subcritical dynamos in shear flows. {\em Astron. Nashr.\/} {\bf 329},
  750--761.

\bibitem[Riols {\em et~al.\/}(2013)Riols, Rincon, Cossu, Lesur, Longaretti,
  Ogilvie \& Herault]{Riols2013}
{\sc Riols, A., Rincon, F., Cossu, C., Lesur, G., Longaretti, {P.-Y.}, Ogilvie,
  G.I. \& Herault, J.} 2013 Global bifurcations to subcritical
  magnetorotational dynamo action in {K}eplerian shear flow. {\em J. Fluid
  Mech.\/} {\bf 731}, 1--45.

\bibitem[{Riols} {\em et~al.\/}(2015){Riols}, {Rincon}, {Cossu}, {Lesur},
  {Ogilvie} \& {Longaretti}]{Riols2015}
{\sc {Riols}, A., {Rincon}, F., {Cossu}, C., {Lesur}, G., {Ogilvie}, G.~I. \&
  {Longaretti}, P.} 2015 {Dissipative effects on the sustainment of a
  magnetorotational dynamo in {K}eplerian shear flow}. {\em Astronomy \&
  Astrophysics\/} {\bf 575}, A14--1--A14--7.

\bibitem[Schmid \& Henningson(2001)]{Schmid2001}
{\sc Schmid, P.~J. \& Henningson, D.~S.} 2001 {\em Stability and Transition in
  Shear Flows\/}. New York: Springer.

\bibitem[Schneider {\em et~al.\/}(2008)Schneider, Gibson, Lagha, De~Lillo \&
  Eckhardt]{Schneider2008}
{\sc Schneider, T.M., Gibson, J.F., Lagha, M., De~Lillo, F. \& Eckhardt, B.}
  2008 {Laminar-turbulent boundary in plane {C}ouette flow}. {\em Phys. Rev.
  E\/} {\bf 78}, 37301.

\bibitem[Schneider {\em et~al.\/}(2007)Schneider, Eckhardt \&
  Vollmer]{Schneider2007}
{\sc Schneider, T.~M., Eckhardt, B. \& Vollmer, J.} 2007 Statistical analysis
  of coherent structures in transitional pipe flow. {\em Phys. Rev. E\/} {\bf
  75}, 066313.

\bibitem[Schoppa \& Hussain(2002)]{Schoppa2002}
{\sc Schoppa, W. \& Hussain, F.} 2002 Coherent structure generation in
  near-wall turbulence. {\em J. Fluid Mech.\/} {\bf 453}, 57--108.

\bibitem[Smagorinsky(1963)]{Smagorinsky1963}
{\sc Smagorinsky, J.} 1963 General circulation experiments with the primitive
  equations: I. the basic equations. {\em Mon. Weather Rev.\/} {\bf 91},
  99--164.

\bibitem[Smith \& Metzler(1983)]{Smith1983}
{\sc Smith, J.~R. \& Metzler, S.~P.} 1983 The characteristics of low-speed
  streaks in the near-wall region of a turbulent boundary layer. {\em J. Fluid
  Mech.\/} {\bf 129}, 27--54.

\bibitem[Smits {\em et~al.\/}(2011)Smits, McKeon \& Marusic]{Smits2011}
{\sc Smits, A.~J., McKeon, B.~J. \& Marusic, I.} 2011 High-reynolds number wall
  turbulence. {\em Ann. Rev. Fluid Mech.\/} {\bf 43}, 353--375.

\bibitem[Tillmark \& Alfredsson(1994)]{Tillmark1994}
{\sc Tillmark, N. \& Alfredsson, H.} 1994 Structures in turbulent plane
  {C}ouette flow obtained from correlation measurements. In {\em Advances in
  Turbulences V\/} (ed. R.~Benzi), pp. 502--507. Kluwer.

\bibitem[Toh \& Itano(2003)]{Toh2003}
{\sc Toh, S. \& Itano, T.} 2003 A periodic-like solution in channel flow. {\em
  J. Fluid Mech.\/} {\bf 481}, 67--76.

\bibitem[Toh \& Itano(2005)]{Toh2005}
{\sc Toh, S. \& Itano, T.} 2005 Interaction between a large-scale structure and
  near-wall structures in channel flow. {\em J. Fluid Mech.\/} {\bf 524},
  249--262.

\bibitem[Tomkins \& Adrian(2003)]{Tomkins2003}
{\sc Tomkins, C.~D. \& Adrian, R.~J.} 2003 {Spanwise structure and scale growth
  in turbulent boundary layers}. {\em J. Fluid Mech.\/} {\bf 490}, 37--74.

\bibitem[Tomkins \& Adrian(2005)]{Tomkins2005}
{\sc Tomkins, C.~D. \& Adrian, R.~J.} 2005 {Energetic spanwise modes in the
  logarithmic layer of a turbulent boundary layer}. {\em J. Fluid Mech.\/} {\bf
  545}, 141--162.

\bibitem[Tsukahara {\em et~al.\/}(2007)Tsukahara, , Iwamoto \&
  Kawamura]{Tsukahara2007}
{\sc Tsukahara, T., , Iwamoto, K. \& Kawamura, H.} 2007 {POD} analysis of
  large-scale structures through {{DNS}} of turbulence {C}ouette flow. In {\em
  Advances in Turbulence XI.\/}, pp. 245--247. Springer.

\bibitem[Tsukahara {\em et~al.\/}(2006)Tsukahara, Kawamura \&
  Shingai]{Tsukahara2006}
{\sc Tsukahara, T., Kawamura, H. \& Shingai, K.} 2006 {DNS} of turbulent
  {C}ouette flow with emphasis on the large-scale structure in the core region.
  {\em J.~Turbulence.\/} {\bf 42}.

\bibitem[Viswanath(2007)]{Viswanath2007}
{\sc Viswanath, D.} 2007 The dynamics of transition to turbulence in plane
  {C}ouette flow. {\em ArXiv physics/0701337\/} .

\bibitem[Waleffe(1995)]{Waleffe1995}
{\sc Waleffe, F.} 1995 Hydrodynamic stability and turbulence: Beyond transients
  to a self-sustaining process. {\em Stud. Appl. Math.\/} {\bf 95}, 319--343.

\bibitem[Waleffe(1998)]{Waleffe1998}
{\sc Waleffe, F.} 1998 Three-dimensional coherent states in plane shear flows.
  {\em Phys. Rev. Lett.\/} {\bf 81}, 4140--4143.

\bibitem[Waleffe(2003)]{Waleffe2003}
{\sc Waleffe, F.} 2003 {Homotopy of exact coherent structures in plane shear
  flows}. {\em Phys. Fluids\/} {\bf 15}, 1517--1534.

\bibitem[Willis {\em et~al.\/}(2010)Willis, Hwang \& Cossu]{Willis2010}
{\sc Willis, A.~P., Hwang, Y. \& Cossu, C.} 2010 Optimally amplified
  large-scale streaks and drag reduction in the turbulent pipe flow. {\em Phys.
  Rev. E\/} {\bf 82}, 036321.

\bibitem[Zhou {\em et~al.\/}(1999)Zhou, Adrian, Balachandar \&
  Kendall]{Zhou1999}
{\sc Zhou, J., Adrian, R.~J., Balachandar, S. \& Kendall, T.~M.} 1999
  Mechanisms for generating coherent packets of hairpin vortices in channel
  flow. {\em J. Fluid. Mech.\/} {\bf 387}, 353--396.

\end{thebibliography}

\newcommand{\noopsort}[1]{} \newcommand{\printfirst}[2]{#1}
  \newcommand{\singleletter}[1]{#1} \newcommand{\switchargs}[2]{#2#1}

%%%%%%%%%%%%%%%%%%%%%%%%%%%%%%%%%%%%%%%%%%%%%%%%%%%%%%%%%%%%%%%%%%%%%%%%%%%%%%%
%%%%%%%%%%%%%%%%%%%%%%%%%%%%%%%%%%%%%%%%%%%%%%%%%%%%%%%%%%%%%%%%%%%%%%%%%%%%%%%
\end{document}